\documentstyle[graphicx,fleqn,twoside,12pt,epsf]{article}
\begin{document}
%\rightline{Unile-COBRAs-00-1}

\newcommand{\bev}{\begin{verbatim}}
\newcommand{\beq}{\begin{equation}}
\newcommand{\beqa}{\begin{eqnarray}}
\newcommand{\beqn}{\begin{eqnarray}}
\newcommand{\eeqn}{\end{eqnarray}}
\newcommand{\eeqa}{\end{eqnarray}}
\newcommand{\eeq}{\end{equation}}
\newcommand{\Eev}{\end{verbatim}}
\newcommand{\bec}{\begin{center}}
\newcommand{\eec}{\end{center}}

\pagestyle{myheadings}
\medskip
\medskip
\centerline{\bf Supersymmetric Scaling Violations (I).}
\centerline{\bf Solving the Supersymmetric DGLAP Evolution}

\medskip
\medskip

\medskip

\medskip

\centerline{ Claudio Corian\`{o}}
\medskip
\centerline{\em Dipartimento di Fisica\footnote{Email: Claudio.Coriano@le.infn.it}}
\centerline{\em Universita' di Lecce}
\centerline{and}
\centerline{\em Istituto Nazionale di Fisica Nucleare}
\centerline{\em Sezione di Lecce}
\centerline{\em Via Arnesano, 73100 Lecce, Italy}

\bigskip
\vspace{.7cm}

\medskip

\begin{abstract}
We analize the renormalization group equations of supersymmetric QCD with $N=1$ 
for the evolution of parton distributions. For this purpose we develope 
a simple recursive algorithm in x-space to include both regular regions
and supersymmetric regions in the evolution in the step approximation.
Supersymmetric distributions are generated within a radiative model, with vanishing 
initial conditions for the superpartners. 
Here we focus on a scenario with broken susy, 
characterized by a lighter gluino coupled to the standard evolution
and a decoupled scalar quark. Predictions for the all the distributions are presented.

\end{abstract}
\newpage

\section{Introduction}
The study of the QCD scaling violations is an important chapter of high energy physics 
and a very important tool for the analysis of future data at the LHC. 
At large energy and momentum transfers, the underlying quark-gluon dynamics 
is light-cone dominated and controlled by a mechanism of collinear radiative 
emission described by logarithmic corrections to the lowest order 
cross section. In this picture, based on the parton model, 
initial quark states are assumed to be massless and the running of the coupling is linked to the number of flavours $n_f$ included in the evolution. Crossing 
intermediate thresholds opens up new channels and new dynamics. 
There is a widely used formalism of perturbative QCD that we think is worth to extend to the supersymmetric case and which might be useful for experimental searches of supersymmetry 
at the LHC. The picture is particularly 
appealing if a ``supersymmetric content `` of the proton is found. However, predictions tied to 
this picture may be used to rule out a possible light gluino and/or a light squark. Other applications involve the study of nucleon collisions with the atmosphere 
as in Ultra High Energy Cosmic Rays, where the center of mass energy 
of the primaries can reach several hundreds of TeV's.  

This formalism involves the study of scaling violations induced on the initial state by the opening of supersymmetric channels prior to reaching the hard scattering phase. 
The analysis  requires the notion of supersymmetric parton distributions, supersymmetric 
factorization formulas for supersymmetric  DIS 
and (susy) hadron-hadron collisions on which we elaborate in detail. 
We remark that supersymmetric versions of parton distributions are easy
to define, by a natural generalization of the usual DIS approach.
The phenomenological validity 
of these extensions are clearly linked to the possibility of detecting scaling violations 
in precision measurements of  final states in p-{$\bar{p}$} collisions, 
for instance in Drell Yan, or in other processes with 
a distinct final state. With this analysis in perspective we start developing accurate tools that we will use in a separate work 
for a quantification of rates for gluino/squarks effects at the LHC. 
Although there is no evidence of supersymmetry at current energies, or of supersymmetry coming 
from the initial state, quantifying with accuracy these effects we believe, 
is still an interesting task. These studies are also of direct theoretical relevance, 
since they tell us more specifically how to merge the usual parton model dynamics with the 
new elementary states that supersymmetry predicts. 

In this paper we start analizing these issues and make a first step toward quantifying the impact of supersymmetry in the initial state. Our analysis, in this paper, 
is focused on the supersymmetric DGLAP evolution, 
which is more involved compared to ordinary QCD. 
We solve the equations using a recursive algorithm that we formulate and that we have tested 
which allows to perform a direct match 
between the various regions of the evolution as we increase the factorization scale and allow for new (supersymmetric) channels. Part of the analysis is of technical nature and several 
sections deal with the implementation of the method. We then study in 2 final sections the implications of the susy evolution and compare it to the standard QCD one. Other aspects of the evolution, 
including some applications for collider processes will be analized in a companion paper. 
    
\section{The Method}
Before moving to Supersymmetric QCD (SQCD), we briefly illustrate the method as it applies to the case 
of ordinary QCD. This simpler case will help us 
establish notations that will be later extended to the supersymmetric evolution. 
We build on previous work of Rossi \cite{Rossi}, 
which has been used before in the previous literature \cite{GordonandGordon, Storrow}, 
although never fully documented in its code implementation. 
We combine that method, originally suggested for QCD, 
with recent approaches that use an analytic evaluation of the convolution integrals (``weights'') 
\cite{Botje}, and generalize it to SQCD.
A recently application to QCD in next-to-leading order (NLO) of this approach has been implemented 
by Chuvakin and Smith \cite{ChuvakinSmith} in their analysis of an evolution 
scheme with a varying number of flavours. To make the discussion self-contained, we have 
elaborated in some detail on the method in an appendix.  

The two-loop running of the coupling constant is defined by 

\beq
{\alpha(Q_0^2)\over 2 \pi}={2\over \beta_0}
{ 1\over \ln (Q^2/\Lambda^2)}\left( 1 - {\beta_1\over \beta_0}
{\ln\ln(Q^2/\Lambda^2)\over \ln(Q^2/ \Lambda^2)} +
O({1\over\ln^2(Q^2/\Lambda^2) })\right)
\eeq

where

\beqa
&& \beta_0={11\over 3} C_G - {4\over 3}T_R n_f \nonumber \\
&& \beta_1={34\over 3} C_G^2 - {10\over 3} C_G n_f - 2 C_F n_f,
\eeqa
where 

\beq
C_G= N, \,\,\,\,\,\,\, C_F={ N^2-1\over 2 N},\,\,\,\,\,\,\,\, T_R={1\over 2}
\eeq

and N is the number of colours while $n_f$ is the number of flavours. 

The solution for the running coupling is given by 
\beqa
\alpha(t)&=&{\alpha(0)\over 2 \pi} e^{-\beta_0/2 t}
\eeqa
with $\alpha(Q_0^2)\equiv \alpha(0) $, and $Q_0$ denoting the 
initial scale at which the evolution starts. 
The evolution equations are of the form 

\beqa  
 Q^2 {d\over d Q^2} {q_i}^{(-)}(x, Q^2) &=& 
{\alpha(Q^2)\over 2 \pi} P_{(-)}(x, \alpha(Q^2))\otimes q_{i}^{(-)}(x, Q^2)
\nonumber \\
 Q^2 {d\over d Q^2}\chi_i(x,Q^2)
&=& 
{\alpha(Q^2)\over 2 \pi} P_{(-)}(x, \alpha(Q^2))\otimes \chi_i(x,Q^2),
\nonumber \\
\eeqa
with 
\beq
\chi_i(x,Q^2)={q_i}^{(+)}(x, Q^2) -{1\over n_F} 
q^{(+)}(x, Q^2)
\eeq
for the non-singlet distributions and 
\beqa
&& Q^2 { d\over d Q^2} \left( \begin{array}{c}
q^{(+)}(x, Q^2) \\
G(x, Q^2) \end{array}\right)=
\left(\begin{array}{cc}
P_{qq}(x,Q^2) & P_{qg}(x,Q^2) \\
P_{gq}(x,Q^2) & P_{gg}(x,Q^2)
\end{array} \right)\otimes 
\left( \begin{array}{c}
q^{(+)}(x, Q^2) \\
G(x,Q^2) \end{array}\right) \nonumber \\
\eeqa
for the singlet sector. 

We have defined, as usual 

\beq
q_i^{(-)}=q_i -\bar{q}_i, \,\,\,\,\,
q^{(+)}_i= q_i + \bar{q}_i, \,\,\,\,\,\,
q^{(+)}\equiv\Sigma= \sum_{i=1}^{n_f} q^{(+)}_i.
\eeq
We introduce the evolution variable
\beq
t= -{2\over \beta_0}\ln {\alpha(Q^2)\over \alpha(Q_0^2)}
\eeq
which replaces $Q^2$. 
The evolution equations are then rewritten in the form 
\beqa
 {d\over dt} q_i^{(-)}(t,x)&=&\left( P^{(0)}(x) + 
{\alpha (t)\over 2 \pi}R_{(-)}(x) +...\right)\otimes q_i^{(-)}(t,x)
\label{one} \\
 Q^2 {d\over d t}\chi_i(x, Q^2)
&=& 
\left( P^{(0)}(x) +
{\alpha(t)\over 2 \pi}R_{(+)}(x)\right)\otimes
\chi_i(x,Q^2),
\label{two} \\
{d\over d t}\left( \begin{array}{c}q^{(+)}(t,x) \\
G(x,t)
\end{array} \right)&=& \left( P^{(0)}(x) +
{\alpha(t)\over 2 \pi} R(x) +...\right)\otimes 
\left( \begin{array}{c}
q^{(+)}(x,t) \\
G(x,t) 
\label{three}
\end{array} \right).
\eeqa
In the new variable $t$, the kernels of the evolution take the form 

\beqa
 R_{(\pm)}(x)& = &P^{(1)}_{(\pm)}(x) - 
{\beta_1\over 2 \beta_0} P^{(0)}_V(x)
\nonumber \\
 R(x) &=& P^{(1)}(x)-{\beta_1\over 2 \beta_0} P^{(0)}(x).
\eeqa
Equations (\ref{one}) and (\ref{two}) are solved 
independently for the variables $q_i^{(-)}$ and 
$\chi_i$ respectively. Finally, the solution $q^{(+)}$ 
of eq.~(\ref{three}) (or the singlet equation) 
is substitued into $\chi_i$ in order to obtain $q_i^{(+)}$. 

The equations can be written down in terms of two singlet evolution operators 
$E_{\pm}(t,x)$ and initial conditions 
$\tilde{q}_{\pm}(x,t=0)\equiv \tilde{q}_{\pm}(x)$ as 

\beq
{d\over dt} E_{\pm}=P_{\pm}\otimes E_{\pm},
\eeq
whose solutions are given by
\beqa
&& q_i^{(-)}(t,x)= E_{(-)}\otimes \tilde{q}_i^{(-)} \nonumber \\
&& \chi_i(t,x)=E_{(+)}\otimes \tilde\chi_i(x).
\eeqa
The singlet evolution for the matrix operator $E(x,t)$

\beqa
&& \left( \begin{array}{cc}
E_{FF}(x,t) & E_{FG}(x,t) \\
E_{GF}(x,t) & E_{GG}(x,t) 
\end{array} \right)
\eeqa

\beq
{d E(x,t)\over d t}= P\otimes E(x,t)
\eeq

is solved similarly as 

\beqa
&&\left( \begin{array}{c}
 q^{(+)}
(t,x)\\
G(t,x)
\end{array}\right)= E(t,x)\otimes 
\left( \begin{array}{c}
\tilde{q}^{(+)}(x)\\
\tilde{G}(x)
\end{array}\right). \\
\nonumber
\label{solution}
\eeqa

The unpolarized leading order kernels are expanded in 
$\alpha$ as  

\beqn
P_{ij}(x,\alpha_s)= 
\left( {\alpha_s\over 2 \pi}\right)P^{(0)}_{ij}(x) + 
\left( {\alpha_s\over 2 \pi}\right)^2 P^{(1)}_{ij}(x)
+ ...
\eeqn

Their LO expressions are given by 
\beq
P^{(0)}_{qq, NS}= C_F\left(\frac{1 + x^2}{1-x}\right)_+
\eeq

for the non-singlet sector, and by 

\beqn
&& P^{(0)}_{qq}(x)=P^{(0)}_{qq,NS}\nonumber \\
&& P^{(0)}_{qg}(x)= 2 T_R n_f \left( x^2 + (1-x)^2\right)\nonumber \\
&& P^{(0)}_{gq}(x)=C_F {1 + (1-x)^2\over x}\nonumber \\
&& P^{(0)}_{gg}(x)= 2 N_c\left( {1\over (1-x)_+} + {1\over x}
-2 + x (1-x)\right) + {\beta_{o}\over 2}\delta(1-x)
\eeqn
in the singlet sector. Some simple identities for the ``plus' $(+)$ distributions  and
their definition are given in the appendix. 

In the actual numerical solution of the equation,
one would like to have at hand a recursion relation which can be
implemented in a computer program dynamically at run-time. 
In order to develope recursion relations we proceed as follow. 
We first observe that as far as we are not
resumming double logarithms in $Q$ and $x$, the solution
eq.~(18) can be expanded as a series of convolution products 
 
\beqa
q(x,t)= q_0(x) +t  P\otimes q_0(x) + \frac{t^2}{2!} P\otimes P\otimes
q_0(x) +...
\eeqa
and transformed into a recursion relation for some coefficients $C_N(x)$
\beqa
C_0(x) &=& \delta(x-1)\nonumber\\
C_{N+1} &=& P\otimes C_N(x)
\eeqa
equivalent to the expansion for the evolution operator
\beq
E(x,t)=\sum_{n=0}^\infty C_N(x) \frac{t^n}{n!}.
\label{transform}
\eeq
Written as a recursion relation, the equation is as easy to implement as
a calculation in moment space, but with no need to perform any moment
inversion.    
The possibility of using this expansion as an
alternative way to evolve parton distributions fast and with accuracy
especially  for larger sets of coupled equations, such as for
supersymmetric theories,
is the subject of the following sections. 
Specifically, Rossi's ansatz \cite{Rossi}, originally introduced in the context of the
photon structure funtion, differs from (\ref{transform}) only by an over all 
normalization $C_N=(-1)^N\left(2/\beta_0 \right)^N$

\beq
{A}_0^{NS}(x) =  \sum_{n=0}^{\infty}{{A}^{NS}_{n}(x)\over n!} 
\log^n\left({\alpha(m_{2 \lambda})\over \alpha(Q_0)}\right). \nonumber  \\
\label{grossi}
\eeq

Connecting this expansion to other expansions is also pretty
straightforward. A very elegant basis in which to expand is the Laguerre
basis, introduced by Furmanski and Petronzio \cite{FP1}. A relation
between Rossi's basis and the FP basis can also be derived but the result 
is not particularly illuminating.

Inserting the expansion (\ref{grossi}) into the evolution equation one 
gets some recursion relations 
for the functions $A_{n+1}(x)$ in terms of the $A_n(x)$. 
These are obtained by comparing left hand side and right hand side of
the evolution equations after equating the logarithmic powers with a
running strong coupling constant $\alpha(Q^2)$. 
A pretty detailed study in the polarized case can be
found in ref.~\cite{GordonandGordon}. 
For computational purposes, the recursion relations for the evaluation
of the $A_n(x)$'s can be implemented in
various ways. The one that we have implemented involves the direct
 solution of the recursion relations as illustrated below.
In leading order in $\alpha(Q^2)$ in QCD we get for the unpolarized kernels

\beqa
\tilde{A}_{n+1}^{V,\pm}(x) & = & \tilde{A}_{n}^{V,\pm}(x)\left[ -\frac{3C_F}{\beta^S_0}- 
\frac{4C_F}{\beta^S_0}\ln (1-x) \right] \nonumber \\
& + & \frac{2 C_F}{\beta^S_0}\int^1_x \frac{dy}{y} (1+z) A_{n}^{V}(y)-
\frac{4 C_F}{\beta^S_0} \int^1_x \frac{dy}{y} 
\frac{y  A_{n}^{V}(y) - x  A_{n}^{V}(x)}{y-x}
\label{expanded}
\end{eqnarray}
int the  non-singlet case and

\begin{eqnarray}
 A_{n+1}^{q+}(x) & = & A_{n}^{q+}(x)\left[ -\frac{3C_F}{\beta_0}- 
\frac{4 C_F}{\beta_0}\ln (1-x) \right] \nonumber \\
& + & \frac{2\, C_F}{\beta_0}\int^1_x \frac{dy}{y} (1+{x\over y}) A_{n}^{q+}(y)
 - {2\over \beta_0} n_f \int_x^1 \frac{dy}{y}\left\{2 \frac{x}{y} {(x-y)\over y} + 1
\right\}A_n^G(y) \nonumber \\
& - & \frac{4 C_F}{\beta_0} \int^1_x \frac{dy}{y} 
\frac{y A_{n}^{q+}(y) - x A_{n}^{q+}(x)}{y-x} 
\label{expanded1}
\end{eqnarray}

\begin{eqnarray}
 A_{n+1}^{G}(x) & = & \frac{-2}{\beta_0}C_F \int_x^1 \frac{dy}{y}
\frac{1 + (1 -z)^2}{z}A_n^{q+}(y)\nonumber \\
& - & \frac{4 N_c}{\beta_0}\int_x^1 \frac{dy}{y}
\frac{y A_n^G(y)- x A_n^G(x)}{y-x} - \frac{4 N_c}{\beta_0}\ln(1-x) A_n^G(x)
\nonumber \\
& - & \frac{4 N_c}{\beta_0}\int_x^1 \frac{dy}{y}\left\{
\frac{1}{z}-2 + z(1-z)\right\}A_n^G(y) - A_n^G(x) \nonumber 
\label{expanded2}
\end{eqnarray}
in the singlet. 
Extensions to the NLO case of this approach are pretty
straightforward, and the procedure will be more clear once we will
discuss its implementation to solve the Supersymmetric DGLAP (SDGLAP) 
equations.

\section{Supersymmetric scaling violations}
In $N=1$ QCD gluons have partners called gluinos (here denoted by $\lambda$)
 and left- and right-handed quarks have complex scalar partners (squarks) which we denote as 
$ \tilde{q_L}$ and $\tilde{q_R}$ with $\tilde{q}=\tilde{q}_L + \tilde{q}_R$ (for left-handed and right-handed squarks 
respectively).

The interaction between the elementary fields are described by the 
$SU(3)$ color gauge invariant and supersymmetric lagrangean 

\beqa
&&{ \cal L}= -{1\over 4} G_{\mu\nu}^a G^{\mu\nu}_a + 
{1\over 2}\bar{\lambda}_a \left( i \not{D}\right) \lambda)_a \nonumber \\
&& + \bar{q}_i I \not{D} q_i + D_\mu \tilde{q}_R D^\mu\tilde{q}^{R}
+ D_\mu \tilde{q}_L D^\mu\tilde{q}^{L} + i g \sqrt{2}
\left(\bar{\lambda^a}_R \tilde{q_{i L}^\dagger } T^a q_{L i} + 
\bar{\lambda^a}_L \tilde{q_{i R}^\dagger } T^a q_{R i}  - {\tt h.\,\,\,c.} \right)
\nonumber \\
&& -{1\over 2} g^2 \left( \tilde{q_{L i}^\dagger} T^a \tilde{q_{L i}}
- \tilde{q_{R i}^\dagger}T^a \tilde{q_{R i}^\dagger}\right)^2 
+ {\tt mass\,\,\,\, terms },\nonumber \\
\eeqa

where $a$ runs over the adjoint of the color group and $i$ denotes the number 
of flavours over which we sum.

Past studies of these effects lead to the conclusion that the information 
that supersymmetric initial states carry along the evolution can be easily absorbed into scaling 
violations coming from ordinary QCD. These studies require the knowledge of the 
matrix of the anomalous dimensions (for all the moments), or of the corresponding Altarelli-Parisi (DGLAP) kernels. 
We recall that the leading 
and next-to-leading anomalous dimensions are known \cite{Antoniadis} since long ago, 
at least in the case of partial supersymmetry breaking, in a scenario characterized 
by a light gluino and a decoupled scalar quark (we present in an appendix the leading order form of these kernels for completeness). 
In this work we consider the possibility of a radiatively generated
gluino distributions, in analogy with the case of standard QCD for the gluons.
In fact, in QCD one can use a simple model of the distributions at low $Q$
$(Q=Q_0)$
and run the DGLAP equations up in energy in order to generate distributions of gluons
at any higher scale $Q_f$. Of course, it is well known that in QCD such initial evolution
scale is model dependent. Various amplitudes at higher energy emerge,
depending on the underlying assumptions on the form of their initial shapes.

This approach has been widely used in the literature and can be 
a way to connect low energy quark models 
-which have no gluons but a phenomenological confining potential - to
{\em true} QCD scattering amplitudes. 
One can assume a zero distribution of glue at the lowest scale, or a model dependent non zero one, 
and then use the evolution equations to dress the matrix elements describing the distributions by 
logarithmic corrections. Once the final scale -here identified as the factorization scale of the process - 
is reached, the distributions are convoluted with the usual -on shell- parton cross sections to generate 
the full hadronic cross section.

 In hadronic collisions, $Q_f$ is usually a fraction of the center-of-mass energy, or a fraction of the large 
$p_T$ of the final state jets. It is not uniquely identified and 
a different choice of $Q_f$ underscores a drastic sensitivity of the expansion on this scale. 

 Alternative choices for density of the constituents at low $Q$ are also possible, but, ultimately, whatever the choice 
for the parton structure, it has to match the scattering data available from 
DIS and colliders experiments. In the case of exact SQCD, 
it is natural to parametrize the parton distributions, now with a susy content in the form 

\beqa
 F_2(x, Q^2) &=& x e_i^2\left(
 q_i(x, Q^2) + \bar{q}_i(x, Q^2) + \tilde{q}_{i R}(x,Q^2) + \bar{\tilde{q}}_{i R}(x, Q^2)\right. \nonumber \\
&&\left. \tilde{q}_{i L}(x,Q^2) + \bar{\tilde{q}}_{i L}(x, Q^2)\right),
\eeqa
where the new elementary constituents (squarks and gluons) can be vanishing at low $Q$ and be radiatively generated by 
the evolution, similarly to the gluon case (in standard QCD) contemplated above. 
In the case of a broken susy, the form of $F_2$ does not change from the
 standard QCD form.  

A gluino or a squark parton distribution (a more detailed analysis will be presented elsewhere) is the exact correspondent 
of a gluon or a quark parton distribution, namely a light-cone dominated diagonal (non local) 
correlation function in spacetime, Fourier transformed to (Bjorken) x-space.    

Similarly to the QCD case, in the case of exact 
$N=1$ supersymmetry we define singlet and non-singlet distributions

\beqa
 q_v(x, Q^2) &=& \sum_{i=1}^{n_f}\left( {q}_i(x,Q^2) - {\bar{q}}_i(x, Q^2)\right) , \nonumber \\
 \tilde{q}_V(x, Q^2) &=& \sum_{i=1}^{n_f}
\left( \tilde{q}_i(x, Q^2) - \tilde{\bar{q}}_i(x, Q^2)\right) \nonumber \\
 q^{+}(x, Q^2) &=& \sum_{i=1}^{n_f} \left({q}_i(x, Q^2) + \bar{q}_i(x, Q^2)\right) \nonumber \\
\tilde{q}^{+}(x, Q^2)&=& \sum_{i=1}^{n_f}\left( \tilde{q}_i(x,Q^2) + \tilde{\bar{q}}_i(x, Q^2)\right).
\eeqa

The evolution equations can be separated in two non-singlet sectors and a singlet one. 
The non-singlet are  

\beqa
Q^2 {d\over d Q^2} q_V(x,Q^2)&=&{\alpha(Q^2)\over 2 \pi} \left( P_{qq}\otimes q_V + P_{q\tilde{q}}\otimes 
\tilde{q_V}\right)  \nonumber \\
Q^2 {d\over d Q^2} \tilde{q}_V(x,Q^2)&=&{\alpha(Q^2)\over 2 \pi} \left( P_{\tilde{q}q}\otimes q_V +P_{\tilde{q}\tilde{q}}\otimes \tilde{q_V}\right),  \nonumber \\
\label{susyns}
\eeqa

and the singlet, which mix $q_V$ and $\tilde{q}_V$ with the gluons and the gluinos are 

\beq
 Q^2 {d\over d Q^2}  \left[\begin{array}{c}   G(x,Q^2)\\ 
\lambda(x,Q^2) \\ q^+(x,Q^2) \\\tilde{q}^+(x,Q^2)  \end{array} \right]
=\left[ \begin{array}{llll} 
         P_{GG} & P_{G \lambda} & P_{G q} &  P_{G \tilde{q}} \\
         P_{\lambda G} & P_{\lambda \lambda} & P_{\lambda q} & P_{\lambda \tilde{q}} \\
         P_{q G} & P_{q \lambda} & P_{q q} & P_{q s} \\
         P_{s G} & P_{s \lambda} & P_{\tilde{q} q} & P_{\tilde{q} \tilde{q}} 
         \end{array} \right] \otimes
  \left[ \begin{array}{c}        
         G(x,Q^2)\\ \lambda(x,Q^2) \\ q^+(x,Q^2) \\ \tilde{q}^+(x,Q^2) \end{array} \right].
\eeq

There are simple ways to calculate the kernel of the SDGLAP evolution by a simple extension of the usual methods. 
The changes are primarily due to color factors. There are also some
basic supersymmetric relations which have to be satisfied that will be
analized below. They are generally broken in the case of decoupling.  
We recall that the supersymmetric version of the $\beta$ function is given at two-loop level by
\beqa
\beta_0^S &=& \frac{1}{3}\left(11 C_A - 2 n_f - 2 n_\lambda \right) 
\nonumber \\
\beta_1^S &=&\frac{1}{3}\left( 34 C_A^2 - 10 C_A n_f - 10 C_A n_\lambda 
- 6 C_F n_F - 6 C_\lambda n_\lambda \right)
\eeqa
with $C_\lambda=C_A=N_c$ for the case of Majorana gluinos 
and the ordinary running of the coupling is replaced by its supersymmetric 
running 

\beq
{\alpha^S(Q_0^2)\over 2 \pi}={2\over \beta_0^S}
{ 1\over \ln (Q^2/\Lambda^2)}\left( 1 - {\beta_1^S\over \beta_0^S}
{\ln\ln(Q^2/\Lambda^2)\over \ln(Q^2/ \Lambda^2)} +
O({1\over\ln^2(Q^2/\Lambda^2) })\right).
\eeq

The kernels are modified both in their coupling $(\alpha \to \alpha^S)$ 
and in their internal structure (Casimirs, color factors, etc.) when moving 
from the QCD case to the SQCD case. In our conventions an index ``S'' stands for a 
supersymmetric component (regular, i.e. non supersymmetric, kernels do not carry such an index), but we will omit it when obvious.

\section{Evolution and Matching}
Susy is necessarily broken in the 
real world and therefore, the way the breaking occurs dictates both the
mass spectrum 
and guides the structure of the QCD scaling violations as well. 
There are several parameters that appear in the evolution, $m_\lambda$ and $m_{\tilde{q}}$, the masses of the gluino and squarks, hence a 
complete analysis includes various scenarios, on which we briefly elaborate. 
In a realistic scenario with a broken susy, the squarks have much larger
mass compared to the quarks and the gluinos have  a Dirac or Majorana mass
$m_\lambda$ (or a combination them).
In our case, in order to establish the evolution scales which are of phenomenological 
interest, it is convenient to assume 1) that the scalar quark decouples from the evolution and 2) that the 2-gluino production threshold 
$m_{2\lambda}= 2 m_{\lambda}$ is the intermediate scale,
separating in $Q^2$ the regular QCD region from the supersymmetric one.  
In this scenario some of the
 splitting functions - specifically $P_{q\lambda}$ and $P_{\lambda q}$ - are zero, since 
no collinear emission is associated with massive partons, unless the symmetry is effectively 
restored. 

We impose the separation condition 
\beq
Q_0 < m_{2 \lambda} < Q
\eeq
and only one non singlet evolution equation is considered 

\beqa
Q^2 {d\over d Q^2} q_V(x,Q^2)&=&{\alpha(Q^2)\over 2 \pi} P_{qq}\otimes q_V.   \nonumber \\
\eeqa

The simplified singlet equations, which mix $q_V$ and $\tilde{q}_V$ with the gluons and the gluinos become 

\beq
 Q^2 {d\over d Q^2}  \left[\begin{array}{c} G(x,Q^2)\\ \lambda(x,Q^2) \\ q(x,Q^2)  \end{array} \right]
=\left[ \begin{array}{lll} 
         P_{GG} & P_{G \lambda} & P_{G q} \\
         P_{\lambda G} & P_{\lambda \lambda} & P_{\lambda q} \\
         P_{q G} & P_{q \lambda} & P_{q q} \\
                  \end{array} \right] \otimes
  \left[ \begin{array}{c}        
         G(x,Q^2)\\ \lambda(x,Q^2) \\ q(x,Q^2)  \end{array} \right].
\eeq

Formally, we can write down the 
expression of the complete solution in the form 

\beqa
q_{NS}(x,Q_f^2) & = & q(x,Q_0^2) + \int_{{Q_0}^2}^{m_{2\lambda}^2}
d\ln Q^2 \,\,P_{NS}(x,\alpha(Q^2)\otimes q_{NS}(x,Q^2) \nonumber \\
 & + & \int_{m_{2\lambda}^2}^{{Q_f}^2} d\ln Q^2 \,\,\,\,\, P_{NS}(x,\alpha(Q^2)) \otimes q_{NS}(x,Q^2)
\nonumber \\
\eeqa
for the non-singlet equation
and 
\beqa
\left[\begin{array}{c} G(x,Q_f^2)\\ \lambda(x,Q_f^2) \\ q(x,Q_f^2)  \end{array} \right]
 & = & \left[\begin{array}{c} G(x,Q_0^2)\\ 0 \\ q(x,Q_0^2)  \end{array} \right]
+\int_{Q^2}^{m_{2\lambda}^2} d\ln Q^2 P(x,\alpha(Q))\otimes 
\left[\begin{array}{c} G(x,Q^2)\\ 0 \\ q(x,Q^2)  \end{array} \right] \nonumber \\
& + & 
\int_{m_{2\lambda}^2}^{Q_f^2} d\ln Q^2 P^S(x,\alpha^S(Q^2))\otimes 
\left[\begin{array}{c} G(x,Q^2)\\ 0 \\ q(x,Q^2)  \end{array} \right]
\eeqa
for the singlet one. In analogy with eq.~(\ref{grossi}),
now we introduce supersymmetric coefficients $\tilde{A}^n(x)$ beside the
usual non supersymmetric $A_n(x)$, and impose recursion relations on the
initial ansatz $q(x,Q_0^2)$ of the form 

\beqa
A^{NS}_0(x) & = & \delta(1-x)\otimes q(x,Q_0^2)=q(x,Q_0^2), \\
A^{NS}_{n+1} & = & -{2\over \beta_O}P_{qq}\otimes A_n,  \\
\tilde{A}_0^{NS}(x) & = & \sum_{n=0}^{\infty}{{A}^{NS}_{n}(x)\over n!} 
\log^n\left({\alpha(m_{2 \lambda})\over \alpha(Q_0)}\right),  \\
\tilde{A}^{NS}_{n+1} & = & -{2\over \beta_O^S}P_{qq}^S\otimes \tilde{A}^{NS}_n \nonumber \\
\eeqa
$n=1,2,...$,
which solve the equation in the non-singlet sector.

In the singlet sector we get the solution

\beqa
\left[\begin{array}{c} G(x,Q_0^2)\\ \lambda(x,Q_0^2) \\ q^+(x,Q_0^2)  \end{array} \right]
& = & \left[\begin{array}{c} G(x,Q_0^2)\\ 0 \\ q(x,Q_0^2)  \end{array} \right]
 +  \sum_{n=1}^{\infty}
\frac{1}{n!}\left[\begin{array}{c}  A_{n}^g(x,Q^2)\\ A_{n}^\lambda(x,Q^2) \\ A_{n}^q(x,Q^2)  \end{array} \right]
\,\,\log^n\left({\alpha(m_{2 \lambda})\over \alpha(Q_0)}\right) \nonumber \\
& + & \sum_{n=1}^{\infty}\frac{1}{n!}
\left[\begin{array}{c} \tilde{A}_{n}^g(x,Q^2)\\ \tilde{A}_{n}^\lambda(x,Q^2) \\ 
\tilde{A}_{n}^q(x,Q^2)  \end{array} \right]
\,\,\log^n\left({\alpha(Q)\over \alpha(m_{2 \lambda})}\right), 
\nonumber \\
\eeqa

constructed recursively through the relations 

\beqa
\left[\begin{array}{c} {A}_{0}^g(x)\\ {A}_{0}^\lambda(x) \\ 
{A}_{0}^q(x)  \end{array} \right]
& = & \left[\begin{array}{c} G(x,Q_0^2)\\ 0 \\ 
q(x,Q_0^2)  \end{array} \right], \\
\left[\begin{array}{c} {A}_{n+1}^g(x,Q^2)\\ {A}_{n+1}^\lambda(x,Q^2) \\ 
{A}_{n+1}^q(x)  \end{array} \right]
& = & -\left(\frac{2}{\beta_0}\right) P \otimes 
\left[\begin{array}{c} {A}_{n}^g(x)\\ {A}_{n}^\lambda(x) \\ 
{A}_{n}^q(x)  \end{array} \right] \\
\left[\begin{array}{c} \tilde{A}_{0}^g(x)\\ \tilde{A}_{0}^\lambda(x) \\ 
\tilde{A}_{0}^q(x)  \end{array} \right]
 & = &
\sum_{n=0}^{\infty}\frac{1}{n!}
\left[\begin{array}{c} \tilde{A}_{n}^g(x)\\ \tilde{A}_{n}^\lambda(x) \\ 
\tilde{A}_{n}^q(x)  \end{array} \right]
\,\,\log^n\left({\alpha(m_{2 \lambda})\over \alpha(Q_0)}\right) \\
\left[\begin{array}{c} \tilde{A}_{n+1}^g(x)\\ \tilde{A}_{n+1}^\lambda(x) \\ 
\tilde{A}_{n+1}^q(x)  \end{array} \right]
& = & -\left(\frac{2}{\beta_0^S}\right) P^S \otimes 
\left[\begin{array}{c} \tilde{A}_{n}^g(x)\\ \tilde{A}_{n}^\lambda(x) \\ 
\tilde{A}_{n}^q(x)  \end{array} \right],\nonumber  \\
\eeqa
$n=1,2,...$, having used a vanishing gluon density at the starting point of the
supersymmetric evolution.

\section{Supersymmetric Relations}
An exact supersymmetric scenario is probably interesting only for
analizing theoretical issues concerning the evolution, or to study
the shape of the distributions at extremely high energies, when, again, 
all the supersymmetric partners effectively become mass degenerate. This
scenario can take place at few
TeV's or at several TeV's, depending upon the assumptions
underlying the way supersymmetry is restored. Here we simply focus our
analysis on a light, up to an intermediate-mass Majorana gluino. Other
aspects of the evolution, such as the interplay of Dirac and Majorana
gluinos and the impact of various patterns of susy breaking will be
considered elsewhere.  

As we have already mentioned, the study of the anomalous
dimensions - in leading order - for $N=1$ QCD can be found in \cite{KounnasRoss}
and now we are going to elaborate on that.  

We recall that in a regularization scheme which is manifestly supersymmetric there are some 
supersymmetric relations which are satisfied by the anomalous dimensions. 
This is true, for instance, in the Dimensional Reduction $\overline{DR}$ 
scheme. 
We recall that to leading order the 
$\overline{MS}$ scheme and the supersymmetric $\overline{DR}$ scheme give coincident results. We recall that in the $\overline{DR}$ scheme the traces are kept in 4 
dimensions, and the chiral projectors are treated as usual, with a completely 
anticommuting $\gamma_5$. The loop momenta are evaluated in n-dimensions. 
In other schemes, such as the t'Hooft-Veltman scheme, $\gamma_5$ is 
instead partially 
anticommuting. The $\overline{MS}$ scheme for chiral states is usually based on this second definition. 
The relation between the two schemes is to leading order 

\beq
P^{(0)}_{\overline{MS}}=P^{(0)}_{\overline{DR}}
\eeq
valid both for the non singlet and the singlet anomalous dimensions. 

The same is true also for the factorization scheme dependence of the coefficient functions. As for the coupling constant, 
the definition of $\alpha$ in the two schemes, the expression of 
$\alpha_{\overline{MS}}$ can be used also in the $\overline{DR}$ scheme as far as the $\Lambda_{QCD}$ scales of the two schemes are related by 
$\Lambda_{\overline{DR}}=
\Lambda_{\overline{MS}}exp\left(C_A/6 \beta_0\right)$. This last change
is tiny and will be neglected. 

Part of the supersymmetric kernels can be obtained at this order 
by a simple change of group factors, from the standard 
QCD kernels. For instance, the substitutions $C_F\to C_\lambda$ 
($C_F=(N_c^2-1)/(2 N_C)$) and $n_f\to n_\lambda$, allow us to obtain the 
kernels $(\lambda\lambda, \lambda g, g \lambda)$ from the ordinary 
kernels $(qq,qg,gq)$. We use 
$C_\lambda= C_A= N_c$, the number of colors. As for the choice of 
the type of representation for the gluino ( Majorana or Dirac), we
recall that in the Dirac 
case we set $n_\lambda= 2 C_A$ while in the Majorana case $n_\lambda = C_A$. 
The organization of the terms in the splitting functions may differ from 
reference to reference, due to the various manipulations one 
can perform on the {\em plus} distributions. 

The supersymmetry relations are given by
\beq
P_{gg} + P_{\lambda g} =  P_{g \lambda} + P_{\lambda \lambda} \\
\label{firstrelation}
\eeq
and by 
\beqa
P_{q g} + P_{\lambda q} & = & P_{g s} + P_{\lambda s}\nonumber \\
P_{q g} + P_{s g}  & = & P_{q \lambda} + P_{s\lambda}\nonumber \\
P_{qq} + P_{s q} & = & P_{q s} + P_{s s}
\eeqa
In general, for a decoupled scalar quark, the only symmetry one would expect 
is eq.~(\ref{firstrelation}). 
In the case of Majorana gluinos, for a scenario with a decoupled 
squark, it is interesting to observe that this relation remains valid
for $x<1$ as well. 
It can be extrapolated to include the $x=1$ point in the case of zero number of flavours 
(supersymmetric gluondynamics).

The evolution has to respect - both in the case of exact susy and of susy 
breaking - 1) baryon number conservation and 2) momentum conservation. 
There are two sum rules associated with these conserved quantities, 
which are generally used to constrain the phenomenological parametrizations 
of the distributions in direct applications. Below the $m_{2\lambda}$
threshold we need to satisfy the usual QCD relations 

\beq
\int_0^1 dx \left(x G(x,Q^2) + x q^{(+)}(x,Q^2)\right)=1
\eeq
for momentum conservation and 
\beq
\int_0^1 dx q^{(-)}(x,Q^2)=3
\label{label3}
\eeq
for baryon number conservation. These require that the anomalous dimensions 
satisfy the relations 
\beqa
\int_0^1 x dx \left(P_{g i}(x) + P_{q i}(x)\right)&=&0 \nonumber \\
\int_0^1 dx P_{NS}(x)&=&0,\nonumber \\
\eeqa
respectively, where $i=q,g$.
In leading order the second equation is simply 
\beq
\int_0^1 dx P^{(0)}_{qq}(x)=0.
\label{label4}
\eeq
Moving above the 2-gluino threshold the momentum sum rule becomes 
\beq
\int_0^1 dx \left(x G(x) + x\lambda(x) +x q^{(+)}(x)\right)=1
\eeq
for momentum conservation, while eq.~(\ref{label3}) remains unaltered. 
We get the new momentum sum rule (or second moment sum rule)
\beqa
\int_0^1 x dx \left(P_{g i}(x) + P_{q i}(x) + P_{\lambda i}\right)&=&0,
\eeqa
with $i=q,g,\lambda$. The supersymmetric version of eq.~(\ref{label4})
is simply obtained by replacing $P^{(0)}_{qq}(x)$ by it supersymmetric
counterpart $P^{S\,(0)}_{qq}(x)$. It can be checked easily that the kernels in the appendix satisy these relations. 

In the case of exact supersymmetry we need to keep into account the
scalar quark contribution in both equations. In particular, 
the equation for the second moment becomes 
\beq
\int_0^1 x\,dx \left(x G(x) + x\lambda(x) +x q^{(+)}(x)+x \tilde{q}^{(+)}(x) \right)=1,
\eeq
which implies that 
 \beqa
\int_0^1 x dx \left(P_{g i}(x) + P_{q i}(x) + P_{\lambda i} +
P_{\tilde{q} i}\right)&=&0,
\eeqa
with $i=q,g,\lambda,\tilde{q}$.
As for baryon number conservation, equation (\ref{label3}) gets modified into 
\beq
\int_0^1 dx \left( q^{(-)}(x) +\tilde{q}^{(-)}(x)\right)=3
\eeq
and using (\ref{susyns}) one gets two relations 
\beqa
\int_{0}^1 dx \left( P^S_{qq} + P^S_{sq}\right)&=&0\nonumber \\ 
\int_{0}^1 dx \left( P^S_{ss} + P^S_{qs}\right)&=&0 \nonumber \\
\eeqa
(below we will drop the susy index $S$ in front of the kernels when obvious).
The first relation in the equation above, for instance, clearly implies that the end point contributions in the ordinary 
$P_{qq}$ kernel are to be modified in order to insure conservation of baryon number. 
In the case of a susy breaking scenario such additional end-point contribution is trivially absent and the form of 
the $P_{qq}$ kernel remains the same, except for the replacement of
$\alpha$ with $\alpha^S$, the supersymmetric coupling and $\beta_0$ with $\beta^S_0$.

\section{ Applications}
In order to apply the formalism to SQCD we proceed by discussing the 
method for a scenario with a broken susy (decoupled squark). 
We proceed from the non singlet case, 
analizing in details the intermediate steps of the algorithm. 

We start with a solution of the {\em standard} non-singlet 
DGLAP equation assuming the boundary condition 
$A_0^V(x)=q^V(x,Q_0^2)$ and run the equation up to the gluino threshold
and construct the coefficients recursively. We arrest the coefficient up
to a desired order $(\bar{n})$, which can be as large as $30$. In
general, the rate of convergence of the asymptotic expansion changes
with the value of the momentum $Q$. 

The solution is then 
constructed in the region $ Q_0 < Q < m_{2 \lambda} $ as 
\beq
q^V(x,m_{2\lambda})=\sum_{n=0}^{\bar{n}}{A^V_n(x)\over n!}\ln\left( 
\frac{\alpha(m_{2\lambda})}{\alpha(Q_0)}\right)
\eeq
and used as initial condition for the next stage of the 
evolution, which involves the region $ m_{2\lambda} < Q < Q_f $, with $Q_f$ 
being the final evolution scale. 

At the next stage, we set $\tilde{A}^V_{0}(x)=\delta(1-x)\otimes q^V(x,m_{2 \lambda})$ and solve recursively 
using the supersymmetric version of the kernels. The strong coupling constant 
$\alpha(Q^2)$ and its running are replaced by their 
supersymmetric version $\alpha^S(Q^2) $, and so are 
the coefficients of the beta function $(\beta_i\to \beta_i^s)$.
Finally the solution is written down in the form 
\beqa
q^{V}(x,Q^2) & = & \sum_{n=0}^{\bar{n}} {A_n^{V}(x)\over n!} 
\log^n\left({\alpha(m_{2 \lambda})\over \alpha(Q_0)}\right)  +  
\sum_{n=1}^{\bar{n}'}{\tilde{A}_n^{NS} \over n!}\log^n\left({\alpha(m_{2 \lambda})\over \alpha(Q)}\right)
\eeqa

where $\bar{n}'$ is the index at which we arrest the supersymmetric recursion. 
In our implementation we have kept the values of $\bar{n}$ and
$\bar{n}'$ very close. 

In the singlet case the procedure is not much different. 
We evolve according to the standard DGLAP equation in the region below 
the 2-gluino threshold, having set to zero 
the contribution from the gluino at the beginning (up to the
supersymmetric threshold).  This 
is equivalent to having set $A_n^\lambda(x)=0$ for any $n$, which means 
that in the region below $m_{2\lambda}$ there is no radiative production of gluinos in the initial stage. We iterate the recursion 
relations up to a given value $\bar{n}$ of the index $n$ with the
standard DGLAP.  
The initial conditions for the next stage of the evolution are then fixed by 
the relations 
\beqa
\tilde{A}^{q+}_0(x) & = & q^+(x,m_{2\lambda}), \nonumber \\
& = & \sum_{n=0}^{\bar{n}}{A^q_n(x)\over n!}\log^n\left( 
\frac{\alpha(m_{2\lambda})}{\alpha(Q_0)}\right),\\
\tilde{A}^{g}_0(x) & = & G(x,m_{2\lambda}) \nonumber \\
& = & \sum_{n=0}^{\bar{n}}{A^g_n(x)\over n!}\log^n\left( 
\frac{\alpha(m_{2\lambda})}{\alpha(Q_0)}\right)\\
\tilde{A}_0^{\lambda}(x)  & = & \lambda(x,Q_0^2) = 0. \nonumber \\
\eeqa
After this, we determine recursively the coefficients $ \tilde{A}_n $  of the supersymmetric expansion 

\beqa
\tilde{A}_{n+1}^{q+} & = & 
-\frac{4 C_F}{\beta_0^S} \int_x^1 \frac{dy}{y}
\frac{ y \tilde{A}_n^{q +}(y) - x \tilde{A}_n^{q +}(x)}{ y - x}\nonumber \\
& - & \frac{4 C_F}{\beta_0^S}\log(1-x) \tilde{A}_n^{q +}(x) + \frac{2 C_F}{\beta_0^S}
\int_x^1 \frac{d y}{y} ( 1 + z) \tilde{A}_n^{q +}(y) -
\frac{3 C_F}{\beta_0^S}\tilde{A}_n^{q +}(x) \nonumber \\
 & - & \frac{2 n_f}{\beta_0^S}\int_x^1 \frac{d y}{y}\left( 1 - 2 z + 2 z^2
\right) \tilde{A}_n^g(y),\nonumber  \\
\eeqa

\beqa
\tilde{A}_{n +1}^\lambda(x) & = & -4 \frac{C_\lambda}{\beta_0^S}\int_x^1\frac{dy}{y}
\frac{ y \tilde{A}_n^\lambda(y) - x \tilde{A}_n^\lambda(x)}{y -x} \nonumber \\
 & + & 2 \frac{C_\lambda}{\beta_0^S}\int_x^1 \frac{dy}{y} ( 1+z) \tilde{A}_n^\lambda(y) - 
\frac{3}{\beta_0^S} C_\lambda\tilde{A}_n^\lambda(x) \nonumber \\
& - & \frac{2}{\beta_0^S} n_\lambda \int_x^1 \frac{dy}{y}(1 - 2 z + 2 z^2)) 
\tilde{A}_n^g(y) - 4 \frac{C_\lambda}{\beta_0^S}\tilde{A}_n^\lambda(x) \log(1-x), \\
\eeqa

\beqa
\tilde{A}_{n+1}^g &=& -\frac{2}{\beta_0^S}C_F \int_x^1 \frac{dy}{y}
\left(\frac{2}{z}-2 + z\right)\tilde{A}_n^q(y)
-\frac{2}{\beta_0^S}C_\lambda\int_x^1 \frac{dy}{y}\left( 
\frac{2}{z}-2 +z\right)\tilde{A}_n^\lambda(y)
\nonumber \\
& & -4\frac{C_A}{\beta_0^S}\int_x^1\frac{dy}{y}\frac{ y \tilde{A}_n^g(y)
- x \tilde{A}_n^g(x)}{y -x} - 4 \frac{C_A}{\beta_0^S}\log(1- x)\tilde{A}_n^g(x)
\nonumber \\
&& - 4\frac{C_A}{\beta_0^S}\int_x^1\frac{dy}{y}\left(\frac{1}{z}-2 + z(1-z)\right)\tilde{A}_n^g(y)
-\tilde{A}_n^g(x).\nonumber \\
\eeqa
Finally, we construct the solution in the form 
\beqa
f(x,Q^2) & = & \sum_{n=0}^{\bar{n}} {A^f_n(x)\over n!} 
\log^n\left({\alpha(m_{2 \lambda})\over \alpha(Q_0)}\right)  +  
\sum_{n=1}^{\bar{n}'}{\tilde{A}_n^f \over n!}\log^n\left({\alpha(Q_f)\over
 \alpha(m_{2\lambda})}\right),\nonumber \\
\eeqa
where we have arrested the supersymmetric recursion up to the index $\bar{n}'$. 
Here $f(x,Q^2)$ indicates a singlet quark, a gluon, or a gluino distribution, with 
$A_n^f(x)$ and $\tilde{A}^f_n(x)$ denoting their corresponding coefficients in the expansion. 
Singularities emerging from the lower integration point $(y=x)$ are the
tricky part of the game, as expected, but can be handled
with various techniques. They will be discussed briefly in the last section.

\section{NLO Extensions}
Let's now move to a next-to-leading order (NLO) analysis of the
evolution. Our discussion here is partial and does not include contributions due to the
emergence of new anomalous dimensions as we move across the
supersymmetric threshold. In recent work \cite{ChuvakinSmith} it has
been shown that in the ordinary distributions of quarks and gluons 
of QCD these effects are
important, especially at small-x. These changes are expected to produce
only a slight modification of the algorithm presented below, and simply
amount to a modification of the boundary condition as we move across the
$m_{2\lambda}$ point. They will not be analized any further in this work
and will be assumed to be negligible. 
As a second point, we remark that the extension of the procedure
outlined below is easy to generalize to the more general case of exact
susy. 

To NLO the ansatz becomes

\beq
q(x,Q^2)=\sum_{n=0}^{\infty} \frac{A_n(x)}{n!}\log^n
\left(\frac{\alpha(Q^2)}{\alpha(Q_0^2)}\right) +
\alpha(Q^2)\sum_{n=0}^\infty \frac{B_n(x)}{n!}\log^n
\left(\frac{\alpha(Q^2)}{\alpha(Q_0^2}\right)
 \eeq
and inserting the usual running of the coupling 

\beqa
\frac{d\alpha}{d \log(Q^2)} & = & \beta(\alpha)
 =  -\frac{\beta_0}{4 \pi}\alpha^2 -\frac{\beta_1}{16 \pi^2} \alpha^3
\eeqa
we get the recursion relations 

\beqa
A_{n+1} & = & -\frac{2}{\beta_0}A_n(x)\nonumber \\
B_{n+1}(x) & = & - B_n(x)- \left(\frac{\beta_1}{4 \beta_0} A_{n+1}(x)\right)
- \frac{1}{4 \pi\beta_0}P^{(1)}\otimes A_n(x) -\frac{2}{\beta_0}P^{(0)}
\otimes B_n(x) \nonumber \\
 & = &  - B_n(x) + \left(\frac{\beta_1}{2 \beta_0^2}P^{(0)}\otimes A_n(x)\right)
S-\frac{1}{4 \pi\beta_0}P^{(1)}\otimes A_n(x) -\frac{2}{\beta_0}P^{(0)}
\otimes B_n(x),  \nonumber \\
\label{recur1}
\eeqa
which are solved with the initial condition $B_0(x)=0$.
The initial condition for the $A_n(x)$  coefficients (i.e. $ A_0(x)$), is specified as in the previous section, with 
$ q(x,Q_0^2) $ identified as the leading order ansatz for the initial
distribution, i.e.
\beq
A_0(x)= \delta(1-x)\otimes q(x,Q_0^2) \equiv q^{(LO)} (x,Q_0^2)
\label{initial}
\eeq

Running the RGE's below the gluino threshold region (up to $Q=m_{2\lambda}$) and solving 
the recursion relations (\ref{recur1}), we get the solution (arrested at a recursive index 
$\bar{n}$)
\beq
q(x,Q^2)=\sum_{n=0}^{\bar{n}}\frac{A_n(x)}{n!}
\log^n\left(\frac{\alpha(Q)}{\alpha(Q_0)}\right) + 
\alpha(Q)\sum_{n=0}^{\bar{n}}\frac{B_n(x)}{n!}
\log^n\left(\frac{\alpha(Q)}{\alpha(Q_0)}\right)
\nonumber \\
\eeq
which is used to fix the intial condition for the second stage 
of the evolution, the supersymmetric one

\beq
q(x,m_{2\lambda})  =  q^{LO}(x,m_{2\lambda}) + \alpha(m_{2\lambda})
\,\,q^{NLO}(x,m_{2\lambda}),
\label{init2}
\eeq
with 

\beqa
q^{LO}(x,m_{2\lambda}) & = & \sum_{n=0}^{\bar{n}}\frac{A_{n}(x)}{n!}
\log^n\left(\frac{\alpha(m_{2\lambda})}{\alpha(Q_0)}\right);\nonumber \\
q^{NLO}(x,m_{2\lambda}) & = & \sum_{n=0}^{\bar{n}}\frac{B_{n}(x)}{n!}
\log^n\left(\frac{\alpha(m_{2\lambda})}{\alpha(Q_0)}\right).\nonumber \\
\eeqa
The supersymmetric recursion relations are then given by 

\beqa
\tilde{A}_0(x) & = & \delta(1-x)\otimes q^{LO}(x,m_{2\lambda}), \nonumber \\
\tilde{B}_0(x) & = & q^{NLO}(x, m_{2\lambda}), \nonumber \\
\tilde{A}_{n+1}(x) & = & -\frac{2}{\beta_0^S}P^{(0)S}\otimes \tilde{A}_n(x), \nonumber \\
\tilde{B}_{n+1}(x) & = & - \tilde{B}_n(x)- \left(\frac{\beta_1^S}{4 \beta_0^S} \tilde{A}_{n+1}(x)\right)
- \frac{1}{4\beta_0^S}P^{(1)S}\otimes \tilde{A}_n(x) -\frac{2}{\beta_0^S}P^{(0)^S}
\otimes \tilde{B}_n(x) \nonumber \\
 & = &  - \tilde{B}_n(x) + \left(\frac{\beta_1^S}{2 {\beta_0^S}^2}P^{(0)S}\otimes 
\tilde{A}_n(x)\right)
- \frac{1}{4\beta_0^S}P^{(1)S}\otimes \tilde{A}_n(x) -\frac{2}{\beta_0^S}P^{(0)S}
\otimes \tilde{B}_n(x).  \nonumber \\
\label{recur2}
\eeqa

We finally construct the general solution in the form 

\beq
q(x,Q^2)= q(x,m_{2\lambda}) + \sum_{n=0}^{\bar{n}}\frac{A_n(x)}{n!}
\log^n\left(\frac{\alpha(Q)}{\alpha(m_{2\lambda})}\right) + 
\alpha(Q)\sum_{n=0}^{\bar{n}}\frac{B_n(x)}{n!}
\log^n\left(\frac{\alpha(Q)}{\alpha(m_{2\lambda})}\right).
\nonumber \\
\eeq
Eq.~(\ref{recur2}) can be expanded in components, since it is valid in matrix form 

\beqa
\tilde{B}_{n+1}^{q+}(x) & = &  - \tilde{B}_n^{q+}(x) + \frac{\beta_1^S}{2 {\beta_0^S}^2}
\left( P^{(0)\,S}_{q\,q}\otimes  \tilde{A}_n^{q+}(x) +
P^{(0)\,S}_{q\,g}\otimes  \tilde{A}_n^{g}(x) +
P^{(0)\,S}_{q \lambda}\otimes  \tilde{A}_n^{\lambda}(x)\right)\nonumber \\
& - & \frac{1}{4\beta_0^S}\left( 
P^{(1)\,S}_{q\,q}\otimes \tilde{A}_n^{q+}(x) +
P^{(1)\,S}_{q\, g}\otimes \tilde{A}_n^{g}(x)+ 
P^{(1)\,S}_{q \lambda}\otimes \tilde{A}_n^{\lambda}(x) \right) \nonumber \\
& - & \frac{2}{\beta_0^S}\left( P^{(0)\,S}_{q\,q}\otimes \tilde{B}_n^{q+}(x)
+ P^{(0)\,S}_{q\, g}\otimes \tilde{B}_n^{g}(x) +
P^{(0)\,S}_{q \lambda}\otimes \tilde{B}_n^{\lambda}(x)\right) \nonumber \\
\eeqa
and similarly for the evolution of the gluon density

\beqa
\tilde{B}_{n+1}^{g}(x) & = &  - \tilde{B}_n^{g}(x) + \frac{\beta_1^S}{2 {\beta_0^S}^2}
\left( P^{(0)\,S}_{g\,q}\otimes  \tilde{A}_n^{q+}(x) +
P^{(0)\,S}_{g\,g}\otimes  \tilde{A}_n^{g}(x) +
P^{(0)\,S}_{g \lambda}\otimes  \tilde{A}_n^{\lambda}(x)\right)\nonumber \\
& - & \frac{1}{4\beta_0^S}\left( 
P^{(1)\,S}_{g\,q}\otimes \tilde{A}_n^{q+}(x) +
P^{(1)\,S}_{g\, g}\otimes \tilde{A}_n^{g}(x)+ 
P^{(1)\,S}_{g \lambda}\otimes \tilde{A}_n^{\lambda}(x) \right) \nonumber \\
& - & \frac{2}{\beta_0^S}\left( P^{(0)\,S}_{g\,q}\otimes \tilde{B}_n^{q+}(x)
+ P^{(0)\,S}_{g\, g}\otimes \tilde{B}_n^{g}(x) +
P^{(0)\,S}_{g \lambda}\otimes \tilde{B}_n^{\lambda}(x)\right) \nonumber \\
\eeqa
while the gluino density is obtained using the recursion relations
\beqa
\tilde{B}_{n+1}^{\lambda}(x) & = &  - \tilde{B}_n^{\lambda}(x) + \frac{\beta_1^S}{2 {\beta_0^S}^2}
\left( P^{(0)\,S}_{\lambda\,q}\otimes  \tilde{A}_n^{q+}(x) +
P^{(0)\,S}_{\lambda\,g}\otimes  \tilde{A}_n^{g}(x) +
P^{(0)\,S}_{\lambda \lambda}\otimes  \tilde{A}_n^{\lambda}(x)\right)\nonumber \\
& - & \frac{1}{4\beta_0^S}\left( 
P^{(1)\,S}_{\lambda\,q}\otimes \tilde{A}_n^{q+}(x) +
P^{(1)\,S}_{\lambda\, g}\otimes \tilde{A}_n^{g}(x)+ 
P^{(1)\,S}_{\lambda\lambda}\otimes \tilde{A}_n^{\lambda}(x) \right) \nonumber \\
& - & \frac{2}{\beta_0^S}\left( P^{(0)\,S}_{\lambda\,q}\otimes \tilde{B}_n^{q+}(x)
+ P^{(0)\,S}_{\lambda\, g}\otimes \tilde{B}_n^{g}(x) +
P^{(0)\,S}_{\lambda\lambda}\otimes \tilde{B}_n^{\lambda}(x)\right). \nonumber \\
\eeqa

Notice that as initial condition for the gluino distributions we take an identically
vanishing function at the scale $Q=m_{2\lambda}$ both in leading and in next-to-leading order.

\beqa
\tilde{A}_0(x) & =  & 0\nonumber \\
\tilde{B}_0(x) & =  & 0.\nonumber \\
\eeqa

\section{Numerical Results} 
It is expected that a large gluino mass, for a fixed factorization scale $Q_f$ in the evolution, 
lowers the size of the scaling violations 
and their impact on the supersymmetric cross section. On the other side, scaling violations induced 
by the susy evolution should grow as we raise the final evolution scale. 
Therefore it seems natural to study the effect of the susy evolution in two different setups  
1) for fixed $m_{\lambda}$ and a varying $Q_f$ or 2) for a varying $m_{\lambda}$ 
at a given factorization scale $Q_f$. We have performed both studies 
and the results are shown  in figs~1-14. 
The implementation of the unpolarized first stage  (QCD) evolution is performed in the $\overline{MS}$ scheme, which is by now standard in most of the high energy physics applications. 
We introduce 
valence quark distributions $q_V(x,Q_0^2)$ and gluon distributions $G(x,Q_0^2)$ at the input scale 
$Q_0$, taken from the 
CTEQ3M parametrization \cite{cteq}
\beqn
q(x)&=&A_0 x^{A_1}(1-x)^{A_2}(1 + A_3 x^{A_4}).
\eeqn
Specifically

\beqn
x u_V(x)&=&1.37x^{0.497}(1-x)^{3.74}[1 + 6.25 x^{0.880}] \nonumber \\
x d_V(x)&=&0.801 x^{0.497}(1-x)^{4.19}[1 + 1.69 x^{0.375}] \nonumber \\
x G(x)&=&0.738 x^{-0.286}(1-x)^{5.31}[1 + 7.30 x] \nonumber \\
x \,\frac{\bar{u}(x) +\bar{d}(x)}{2} &=& 0.547 x^{-0.286}(1-x)^{8.34}[1 +
17.5 x]\nonumber \\
x s(x)&=& 0.5\,x\,\frac{\bar{u}(x) +\bar{d}(x)}{2}\nonumber \\
x\,(\bar{d}-\bar{u})&= & 0.75\, x^{4.97}(1-x)^{8.34}(1 + 30.0\, x)
\eeqa
and a vanishing anti-strange contribution at the input. 
In figs.~1-13, where we have studied the valence quarks, the gluon and the 
gluino distribution for a varying gluino mass ranging from a light to an intermediate gluino 
(10-40 GeV) up to a value of 250 GeV. 
These results generally point toward small scaling violations, which become more 
appreciable as we move closer the smaller-x region (in particular for gluons and gluinos). 
We have chosen as initial evolving scale $Q_i=2$ GeV in all the runs. 

Fig.~1 shows the shape of the distributions close to their initial evolution scale, with 
$Q_i= 2$ GeV and $Q_f=5$ GeV. 
In fig.~2 we show the scaling violations induced by the 
susy evolution on the  $u$ and $d$ quark distributions, 
from the initial scale up to the final scale of 
$100$ GeV for a gluino mass of $10$ GeV. In this figure we show the 
regular versus the supersymmetric evolution for these two distributions. 
The modifications appear to be quite sizeable. The presence of supersymmetry in the evolution 
shows up as a lowering of the maxima of the distributions with a shift toward the small-x region.  

A better distinction 
between the non susy from the susy result is illustrated in fig.~3, which shows a comparison 
between regular and susy evolution for gluons. As expected, these differences get more pronounced moving toward the 
region of smaller $x$, due to the rise of the gluon distribution and to the small-x structure 
of the kernels. 
In fig.~4 we show the singlet $q^{(+)}$ distribution. The difference between the regular and the supersymmetric evolution is 
slim, given the low factorization scale (100 GeV) chosen in this run. 

Gluino and gluon distributions are shown in fig.~5. The gluino distribution is approximately 2 orders of magnitude smaller 
compared to the the supersymmetric 
gluon distribution and grows fast at small-x. Fig.~6 shows a plot of the gluon distribution for a large final factorization scale $Q_f=1$ TeV and a varying gluino mass (100,150 and 200 GeV respectively). Shown is also the regular evolution of the gluon density, 
which is lower than the susy ones at larger x vales and faster growing at smaller x values. 
Fig.~7 shows the factorization scale dependence of the gluino distribution for a sizeable gluino mass (100 GeV) and large 
factorization scales $Q_f=1, 2$ and 5 TeV. 
We plot in fig.~8 the gluon and the gluino distributions for very large evolution scales for a realistic gluino mass of 250 GeV 
and varying factorization scales. We have chosen $Q_f=5$ and 10 TeV respectively. 
The dependence on the final scale appears to be quite mild. Fig.~9 shows the variation of the gluino distribution for different gluino masses (40,100,200 and 250 GeV) and a fixed final evolution scale $Q_f=1$ TeV. 
There is a reduction of the small-x growth of this distribution at smaller-x values 
as the mass of the gluino is raised. In fig.~10 we show the dependence of all the quark distributions on the 
factorization scale $Q_f=5$ and 10 TeV in the case of a supersymmetric scale $m_{2 \lambda}$ of 250 GeV. 

We study the impact of these corrections on future collider experiments by showing results 
for the 2-gluino production in a p-p collision. As can be seen from fig.~11, the production cross section 
is quite small -compared to standard QCD rates-, but gets enhanced by the inclusion of susy 
scaling violations especially at larger energies. We have set the gluino 
mass at 250 GeV. In the same figure we show the dependence of the 
cross section on the factorization scale (510,600 and 700 GeV). The dependence is sizeable, although the 
total rates remain small compared to the QCD background at these energies ( 650 GeV - 2 TeV). 
We have shown in figs.~12 and 13 the partial contributions to fig.~11 coming 
from the $g\, g \to \lambda \lambda$ channel and the $q\, \bar{q}$ channel. The gluon contribution 
is the dominant part in both the regular QCD case and in the supersymmetric case.

\section{Conclusions}
We have solved the supersymmetric DGLAP equation in a scenario with 
a coupled gluino and a decoupled squark using an algorithm based on x-space. 
In particular, 
we have illustrated the evolution of the distributions of quarks and gluons and their supersymmetric 
versions using a radiative model. Although the window on a light gluino 
is now rapidly closing and the hope to detect supersymmetry from scaling violations in the 
initial state at the LHC with a light gluino is slim, the possibility of analizing experimentally the 
impact of heavier supersymmetric particles in the initial state remains, however, an important issue. 
In a specific example (p-p $\to \lambda \lambda$), 
we have seen that the total cross section is sensitive to initial state susy scaling violations and 
on the factorization scale chosen. For 2-gluino production, for instance, 
the rate is small, much below the usual QCD 
background, but gets sizeably enhanced when susy is included. 
Would be interesting to see how much we can rise the 
gluino mass and still obtain a signal on a final state which can not be compensated by 
the usual re-adjustement of the several parameters that describe the ordinary QCD 
parton distributions, or is comparable to it. 
In particular, the strong factorization scale dependence may be reduced 
by the inclusion of radiative corrections in the initial state. 
This and other related issues will be analized elsewhere.

\centerline{\bf Acknowledgements}
I thank R. Pisarski and the Theory Group at Brookhaven
National Laboratory and the C. N. Yang Institute for Theoretical Physics at Stony Brook for hospitality during the final stage of this investigation. 
I am grateful to R. Parwani, J. Smith, G. Sterman, J. Verbaarschot and A. Chuvakin at Stony Brook and to A. Stathopoulos of William Mary for discussions. 
This work is partially supported by MURST and by INFN 
(iniziativa specifica BARI-21).

\subsection{Appendix 1: The Kernels}
Various relations among different types of $``+''$ distributions can be derived. The kernels given in the literature may differ in their final expressions 
due to rearrangements of the corresponding ``+'' functions. 
We give here the expression of these kernels and illustrate some of the 
manipulations needed to reorganize them in a standard form.

In the derivation of the recursion relations we have used the identity 

\beq
\int_x^1 \frac{dy}{y}\left(\frac{1}{(1 - x/y)_+}\right) A_n(y)
= \int_x^1 \frac{dy}{y} {y A_n(y) - x A_n(x)\over y -x} + \log(1-x) A_n(x)
\eeq
Similarly, it is not hard to show the identity of the convolution products 

\beqa
\left({ 1 + x^2\over 1 -x}\right)_+\otimes A_n(x) & = &
\int_x^1\frac{dy}{y}\frac{y^2 + x^2}{(y-x) y}A_n(y) -
A(x)\int_0^x dz \frac{1 + z^2}{1 -z} - A(x)\int_x^1 \frac{dy}{y} 
\frac{ y^2 + x^2}{y^2 ( y-x)} \nonumber \\
\eeqa
with
\beqa
\left[ \left({ 2\over (1-x)_+}\right) 
-1-x +\frac{3}{2}\delta(1-x)\right] \otimes A & = &
\int_x^1 \frac{dy}{y} 2 {y A_n(y) - x A_n(x)\over y -x} \nonumber \\
& - & 2 \log(1-x) A_n(x)
+ \frac{3}{2} A(x) -\int_x^1 \frac{dy}{y}\left(1 + z\right) A_n(y) \nonumber \\
\eeqa
After some manipulations one can show that the two expressions given above 
are equal and therefore
\beq
\left({ 1 + x^2\over 1-x}\right)_+ = {2\over (1-x)_+} -1-x + \frac{3}{2}
\delta(1-x)
\eeq

In leading order, 
the supersymmetric expression of the standard (qq,qg,gq,gg) QCD kernels are obtained by 
replacing $\alpha\to \alpha^S$, and the dependence on the coefficient of the 
$\beta$ function by its supersymmetric version $\beta_0\to \beta_0^S$.

The leading order expression of the kernels in the case of a decoupled scalar quark 
are given below. We have omitted the superscrit ``S'' for simplicity. We remark that  
the usual QCD kernels, after the embedding in the supersymmetric evolution, do not 
acquire new terms at $x=1$, except for $P_{gg}$.  
\beqa
P^{(0)\,}_{gg} & = & 2 C_A\left[ \frac{1}{(1-x)_+} + \frac{1}{x} -2 + 
x(1-x)\right] + \frac{\beta_0^S}{2}\delta(1-x) \nonumber \\
P^{(0) }_{g \lambda} & = & C_\lambda \left[ \frac{1 +(1-x)^2}{x}\right] \nonumber \\
P^{(0) }_{g q} & = & P^{(0)}_{g q}=  C_F\left[ \frac{2}{x}  -2 + x  \right] \nonumber \\
P^{(0)}_{\lambda g}& = & n_\lambda\left[  1 - 2 x + 2 x^2 \right]\nonumber \\
P^{(0)}_{\lambda \lambda} & = & C_\lambda \left[ \frac{2}{(1-x)_+} -1 -x +
\frac{3}{2} \delta(1-x) \right]= C_\lambda\left(\frac{1 + x^2}{(1-x)}
\right)_+ \nonumber \\
P^{(0)}_{\lambda\, \tilde{q}} & = & 0 \nonumber \\
P^{(0)}_{\tilde{q}\,\lambda } & = & 0 \nonumber \\
P^{(0)}_{\tilde{q}\,\tilde{q} } & = & 0 \nonumber \\
P^{(0)}_{\tilde{q}\,q } & = & 0 \nonumber \\
P^{(0)}_{q\,\tilde{q} } & = & 0 \nonumber \\
P^{(0)}_{q g} &=& P^{(0)}_{q g}= n_f\left[ x^2 + (1-x)^2\right] \nonumber \\
P^{(0)}_{q \lambda} & = & n_f(1-x) \nonumber \\
P^{(0)\,S}_{qq} & = & C_F\left[{(1 + x^2)\over (1-x)} \right]_+\nonumber \\ 
 & = & C_F\left( {2\over (1-x)_+} -1-x + \frac{3}{2}\delta(1-x)\right)\nonumber \\
\eeqa

\section{Appendix 2. Discretizations}

For the calculation of the weights used in the numerical analysis we follow closely ref.~\cite{Botje} also implemented in \cite{ChuvakinSmith}. 
At each x-value, these authors use an approximation characterized by 
weights which are calculated analytically, together with an interpolation formula for the 
integration function. The method allows to monitor the singularity appearing in the 
small-x region and to achieve a very good numerical accuracy. The method
speeds up in time the calculation by a large factor, but becomes tedious when
moving to a higher order, since all the integrals have to be {\em exactly}
discretized and the logarithms extracted in each sub-interval.
Here we briefly illustrate the method as it applies to our case.

We briefly recall the numerical strategy 
employed in this analysis. 
We define $\bar{P}(x)\equiv x P(x)$ and $\bar{A}(x)\equiv x A(x)$. 
We also define the convolution product  

\beq
J(x)\equiv\int_x^1 \frac{dy}{y}\left(\frac{x}{y}\right) P\left(\frac{x}{y}\right)\bar{A}(y). \ 
\eeq
The integration interval in $y$ at any fixed x-value is partitioned in an array of 
increasing points ordered from left to right 
$\left(x_0,x_1,x_2,...,x_n,x_{n+1}\right)$ 
with $x_0\equiv x$ and $x_{n+1}\equiv 1$ being the upper edge of the integration 
region. One constructs a rescaled array 
$\left(x,x/x_n,...,x/x_2,x/x_1, 1 \right)$. We define 
$s_i\equiv x/x_i$, and $s_{n+1}=x < s_n < s_{n-1}<... s_1 < s_0=1$.
We get 
\beq
J(x)=\sum_{i=0}^N\int_{x_i}^{x_{i+1}}\frac{dy}{y}
\left(\frac{x}{y}\right) P\left(\frac{x}{y}\right)\bar{A}(y) 
\eeq
At this point we introduce the linear interpolation 
\beq
\bar{A}(y)=\left( 1- \frac{y - x_i}{x_{i+1}- x_i}\right)\bar{A}(x_i) + 
\frac{y - x_i}{x_{i+1}-x_i}\bar{A}(x_{i+1})
\label{inter}
\eeq
and perform the integration on each subinterval with a change of variable $y->x/y$ and replace the integral $J(x)$ with 
its discrete approximation $J_N(x)$
to get 
\beqa
J_N(x) &=& \bar{A}(x_0)\frac{1}{1- s_1}\int_{s_1}^1 \frac{dy}{y}P(y)(y - s_1) \nonumber \\
&+& \sum_{i=1}^{N}\bar{A}(x_i) \frac{s_i}{s_i - s_{i+1}}
\int_{s_{i+1}}^{s_i} \frac{dy}{y}P(y)(y - s_{i+1})\nonumber \\
& -& \sum_{i=1}^{N}\bar{A}(x_i) \frac{s_i}{s_{i-1} - s_{i}}
\int_{s_{i}}^{s_{i-1}} \frac{dy}{y}P(y)(y - s_{i-1}). \nonumber \\
\eeqa
Introducing the coefficients  $W(x,x)$ and $W(x_i,x)$, the integral 
is cast in the form 
\beq
J_N(x)=W(x,x) \bar{A}(x) + \sum_{i=1}^{n} W(x_i,x)\bar{A}(x_i)  
\eeq
where
\beqa
W(x,x) &=& \frac{1}{1-s_1} \int_{s_1}^1 \frac{dy}{y}(y- s_1)P(y), \nonumber \\
W(x_i,x) &=& \frac{s_i}{s_i- s_{i+1}}
\int_{s_{i+1}}^{s_i} \frac{dy}{y}\left( y - s_{i+1}\right) P(y) \nonumber \\
& -& \frac{s_i}{s_{i-1} - s_i}\int_{s_i}^{s_{i-1}}\frac{dy}{y}\left(
y - s_{i-1}\right) P(y).\nonumber \\
\eeqa

We recall that 
\beq
  \int_0^1 dx \frac{f(x)}{(1-x)_+}=\int_0^1 {dy}\frac{f(y)- f(1)}{1-y}
\eeq
and that 
\beq
 \frac{1}{(1-x)_+}\otimes f(x)\equiv 
\int_x^1\frac{dy}{y}\frac{\,\,y f(y) - x f(x)}{y-x} + f(x)\log(1-x) 
\eeq
as can can be shown quite straightforwardly. 

We also introduce the expressions 

\beqa
In_0(x) & = & 
 \frac{s_1}{1- s_1} \log(s_1) + \log(1- s_1) \nonumber \\
\nonumber \\
Jn_i(x) & = & \frac{1}{s_i - s_{i +1}}
\left[ \log\left(\frac{1 - s_{i+1}}{1 - s_i}\right) 
+ s_{i+1} \log\left(\frac{1- s_i}{1 - s_{i+1}}\frac{s_{i+1}}{s_i}\right)\right]
\nonumber \\
Jnt_i(x) & = & \frac{1}{s_{i-1}- s_i}\left[ \log\left(\frac{1 - s_i}{1 - s_{i-1}}\right) 
+ s_{i-1}\log\left( \frac{s_i}{s_{i-1}}\right) + s_{i-1}\left( 
\frac{1 - s_{i-1}}{1 - s_i}\right)\right],   \,\,\,\,\,\ i=2,3,..N \nonumber \\
Jnt_1(x) &=& \frac{1}{1- s_1}\log s_1. \nonumber \\
\eeqa

Using the linear interpolation formula (\ref{inter})  we get the 
relation 

\beqa
\int_x^1\frac{dy}{y} \frac{ y A_n(y) - x A_n(x)}{y-x} &=&
- \log(1-x) A_n(x) + A_n(x) In_0(x)\nonumber \\
&&  + \sum_{i=1}^N A_n(x_i) \left( Jn_i(x) - Jnt_i(x)\right)
\eeqa

which has been used for a fast 
and accurate numerical implementation of the recursion relations.

\newpage

\begin{figure}[thb]
\centerline{\includegraphics[angle=0,width=.8\textwidth]{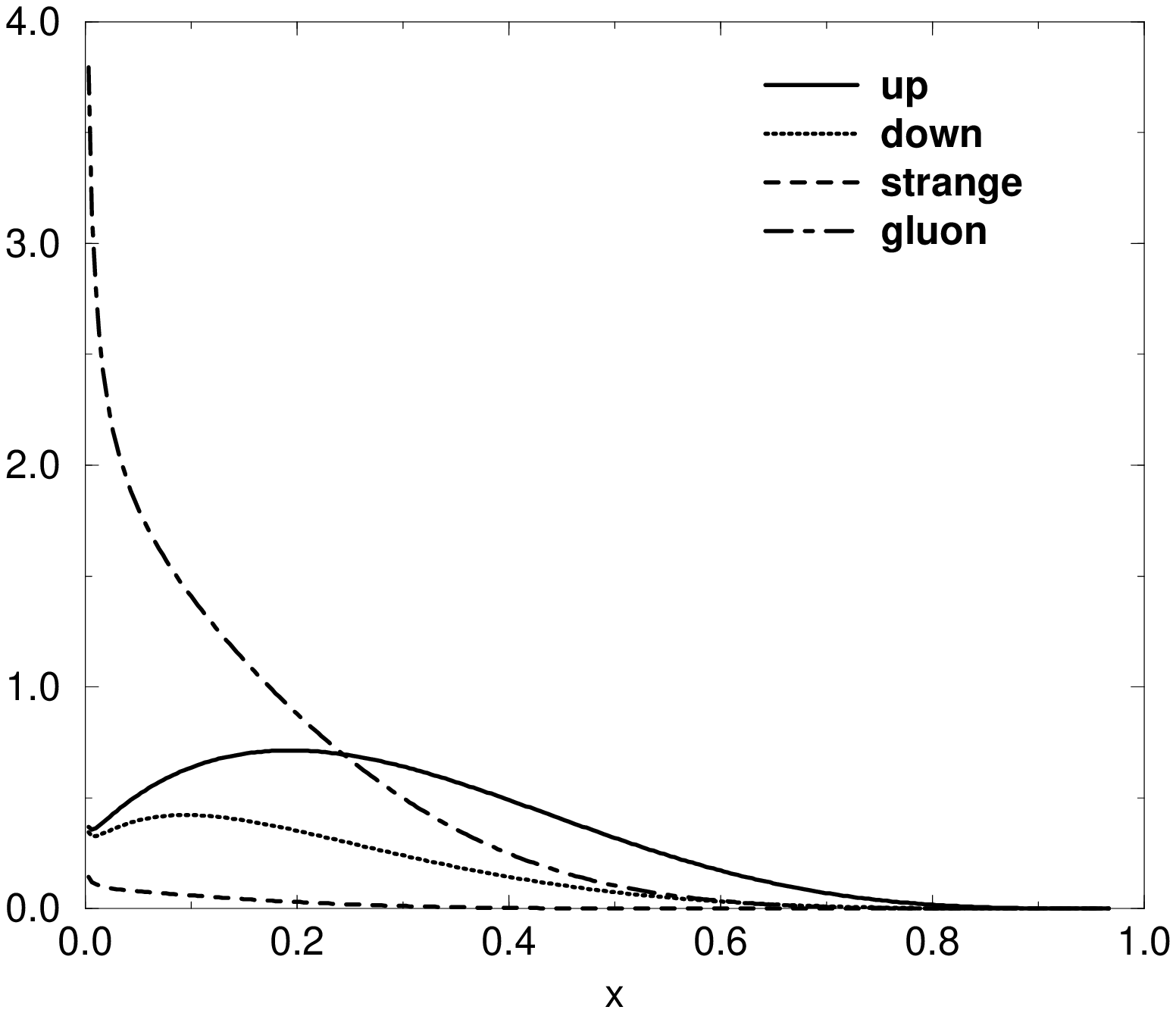}}
\caption{u, d, s and gluon distributions at $Q_i=5$ GeV }
\label{quark0}
\centerline{\includegraphics[angle=0,width=.8\textwidth]{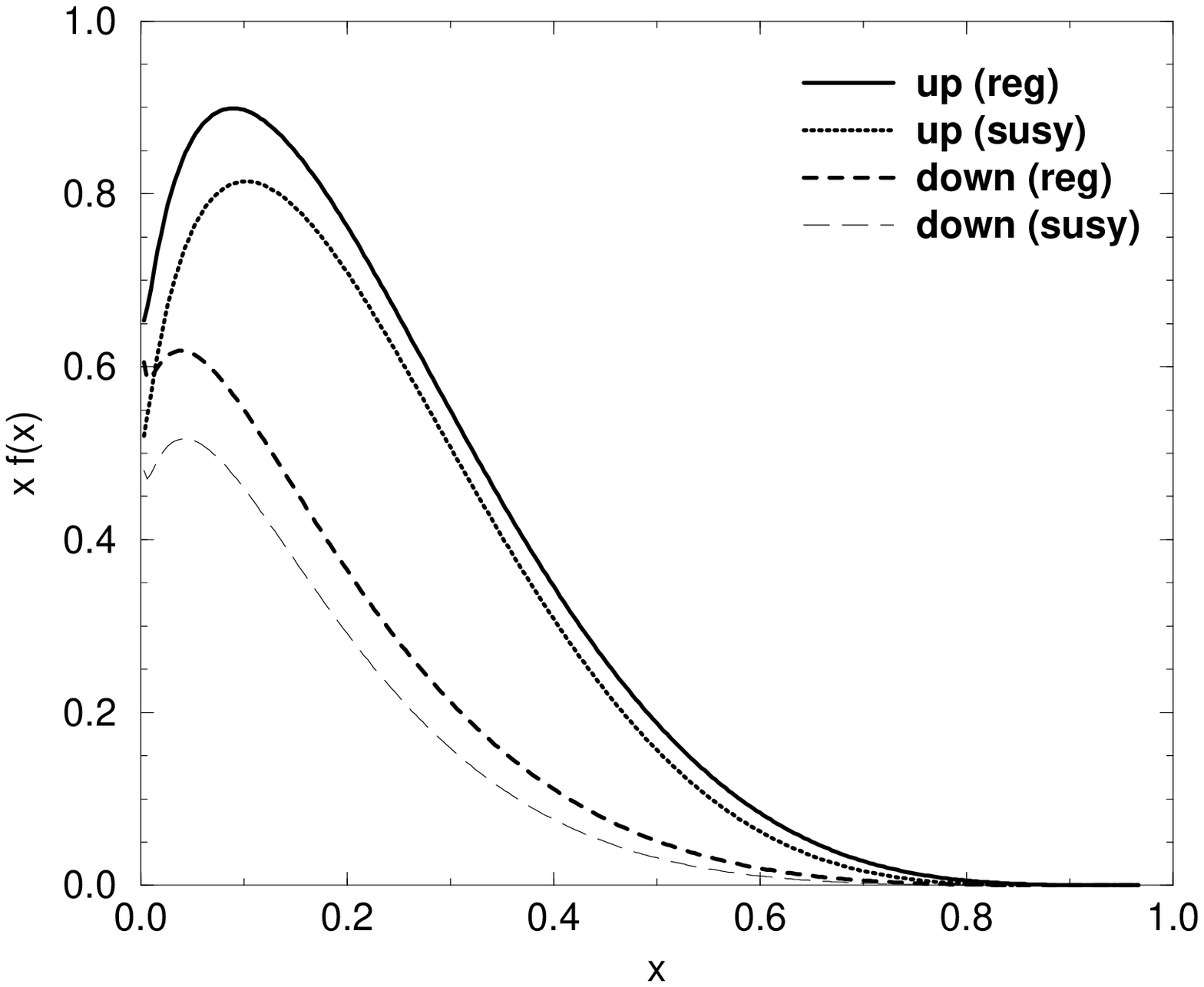}}
\caption{$x u(x)$ and  $x d(x)$ evaluated with $Q_i=2.0$ GeV
 and $Q_f=100$ GeV with $m_{\lambda}=10$ GeV in the standard (reg) and
 susy  evolution }
\label{quark2}
\end{figure}

\begin{figure}
\centerline{\includegraphics[angle=0,width=.8\textwidth]{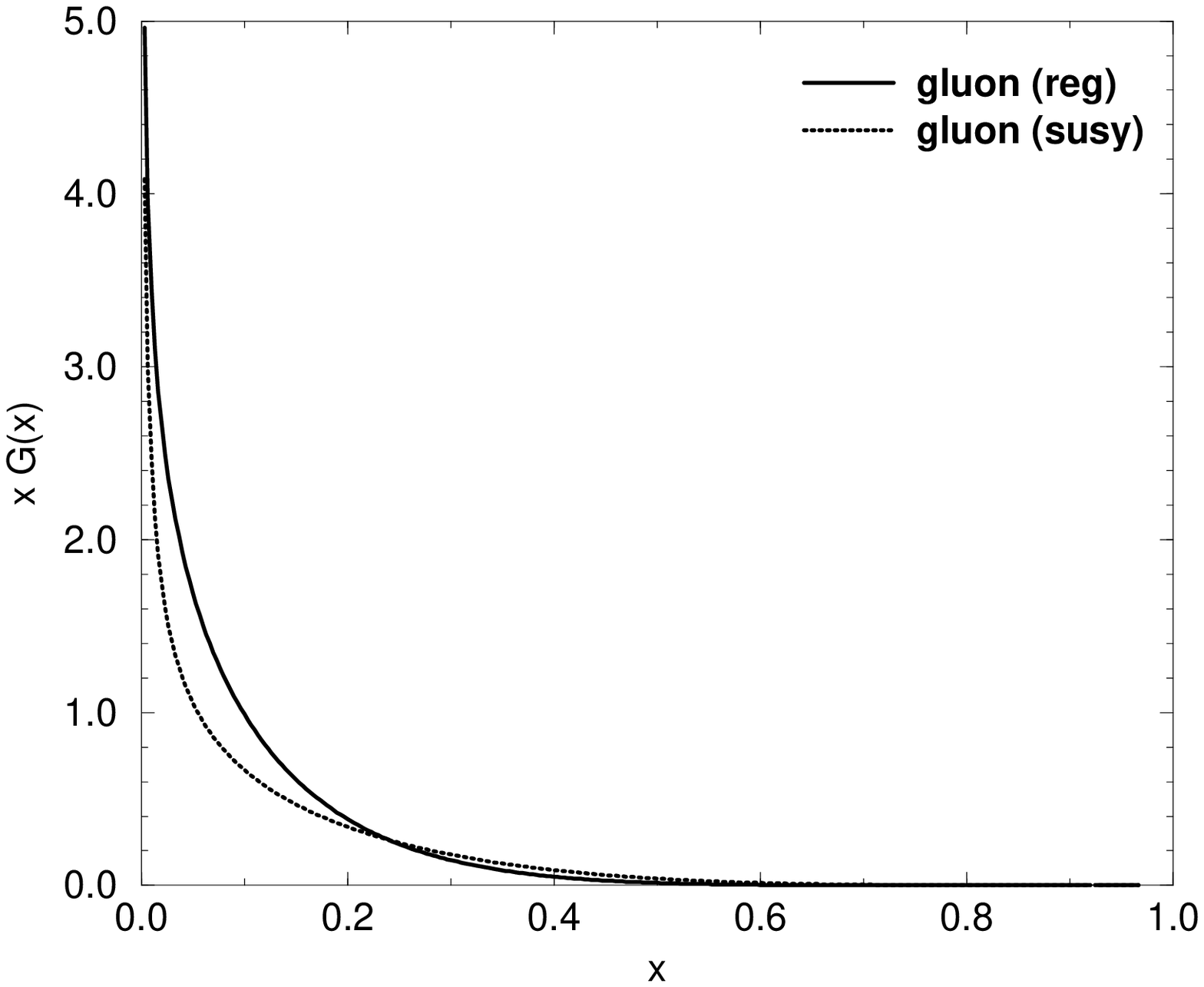}}
\caption{Gluon distributions with $Q_i=2.0$ GeV
and $Q_f=100$ GeV with intermediate $m_\lambda=10$ GeV. The regular and
the susy evolution are shown}
\label{gluon2}
\centerline{\includegraphics[angle=0,width=.8\textwidth]{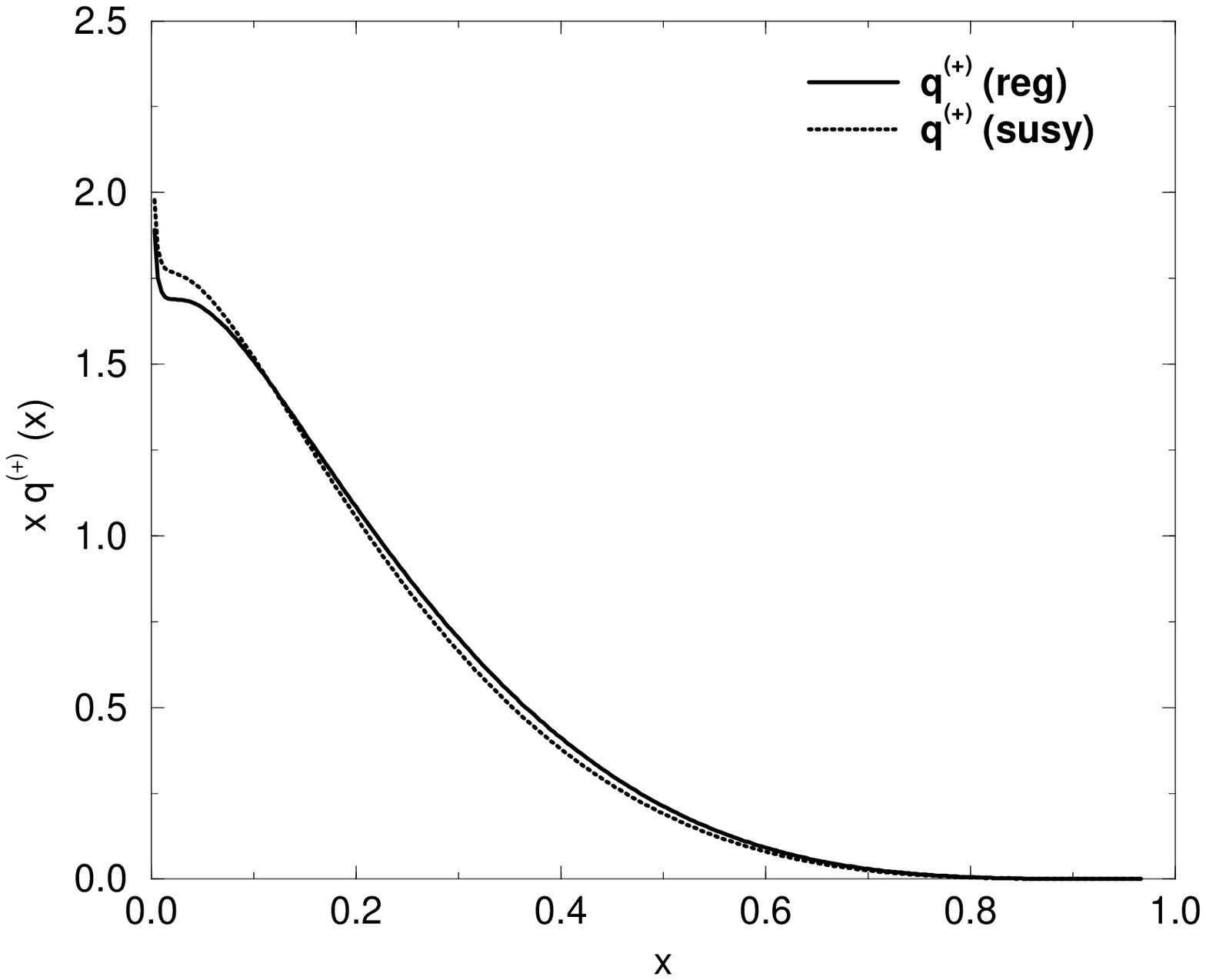}}
\caption{$x q^{(+)}(x)$ (singlet) quark distribution evaluated with $Q_i=5.0$ GeV
and $Q_f=100$ GeV with $m_\lambda=10$ GeV in the standard (non-susy) and
susy evolution}
\label{quark5}
\end{figure}

\begin{figure}
\centerline{\includegraphics[angle=0,width=.8\textwidth]{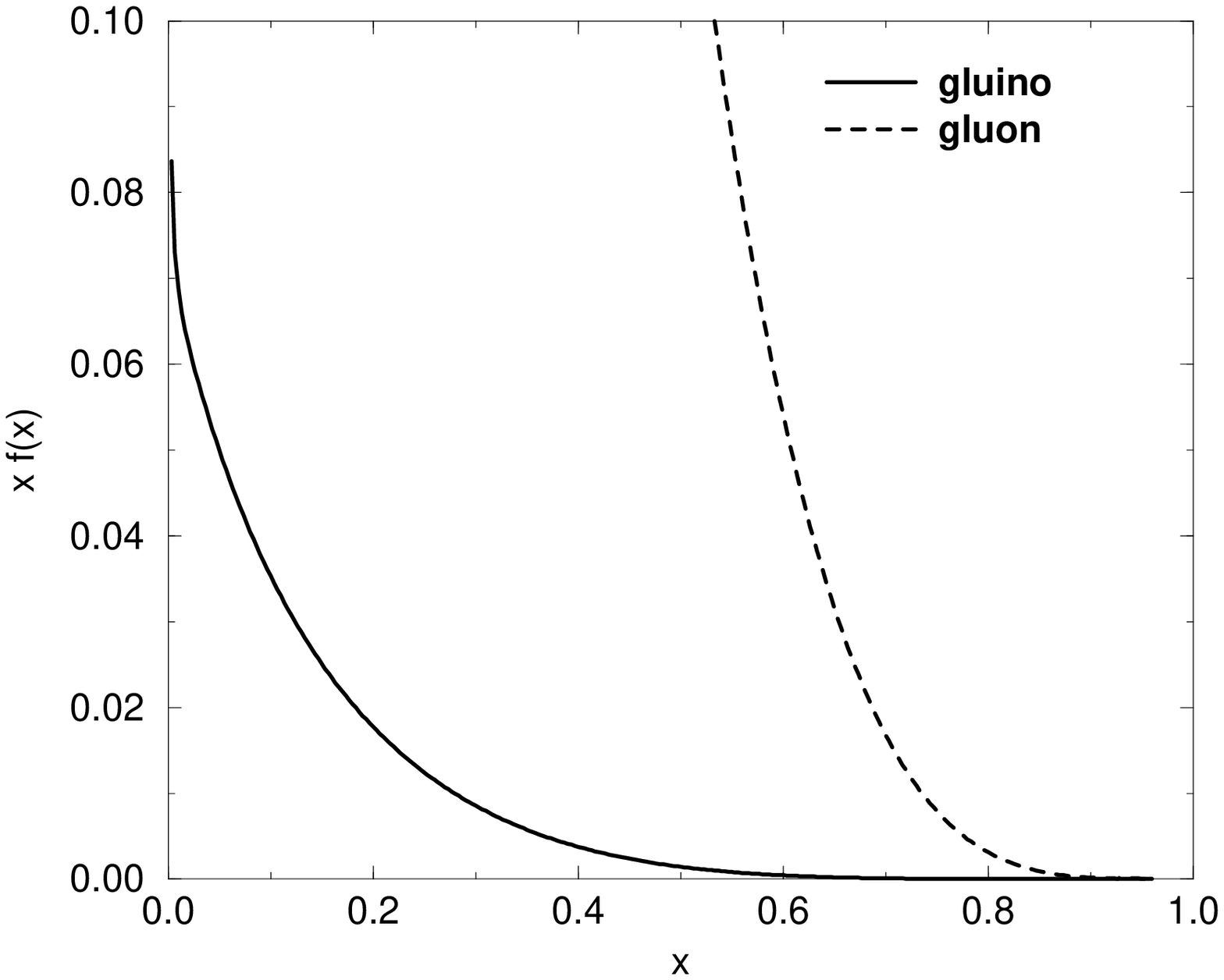}}
\caption{Comparison of the 
supersymmetric gluon and gluino distributions for $m_{\lambda}=30$ GeV and $Q_f= 
100$ GeV.}
\label{singlet5}
\end{figure}

\begin{figure}
\centerline{\includegraphics[angle=0,width=.8\textwidth]{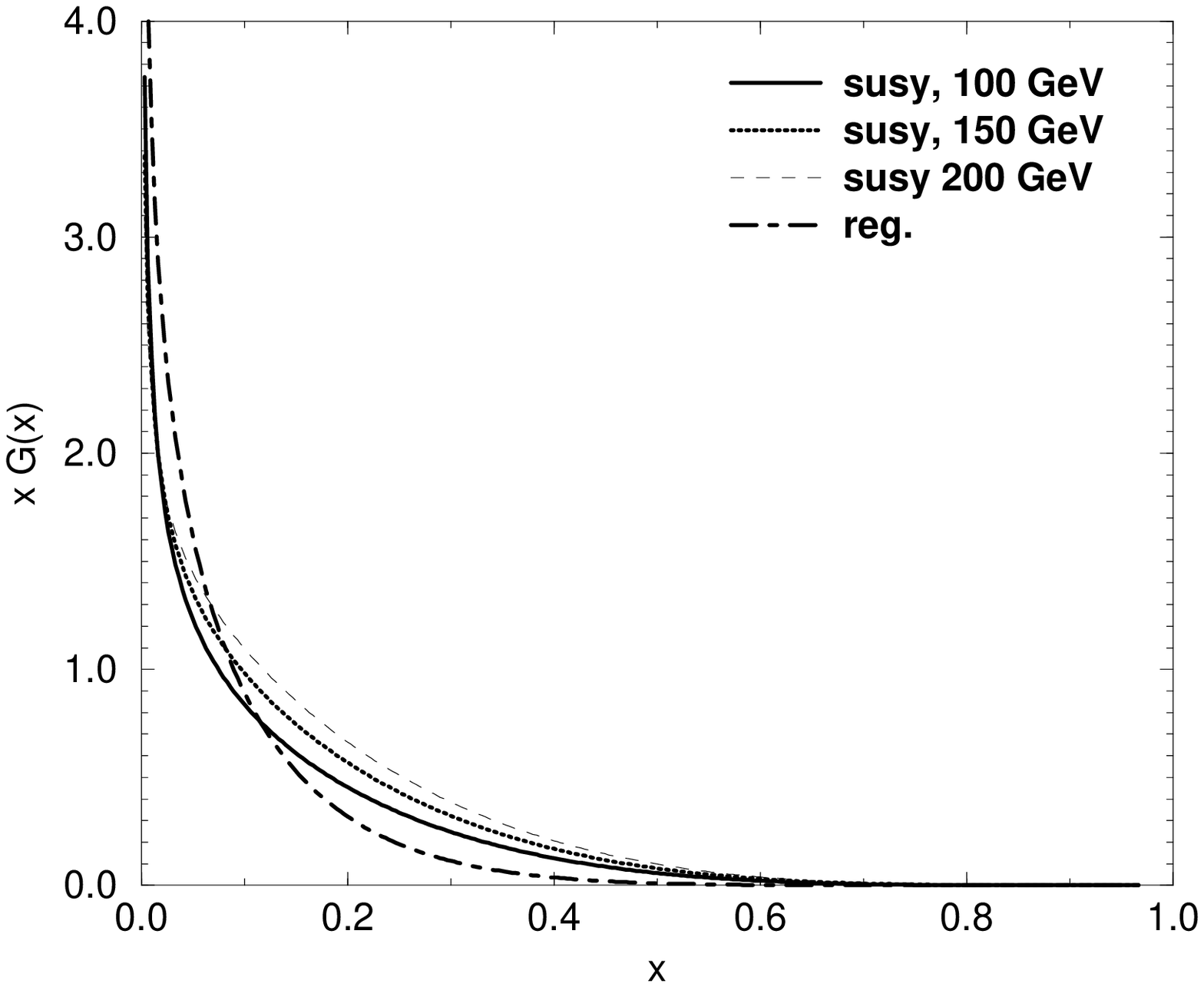}}
\caption{Gluon distribution for 3 values of the gluino mass 100, 150 and 200 GeV 
and a final evolution scale 
$Q_f=1$ TeV. Shown is also the regular (reg) evolution of the same distribution}
\centerline{\includegraphics[angle=0,width=.8\textwidth]{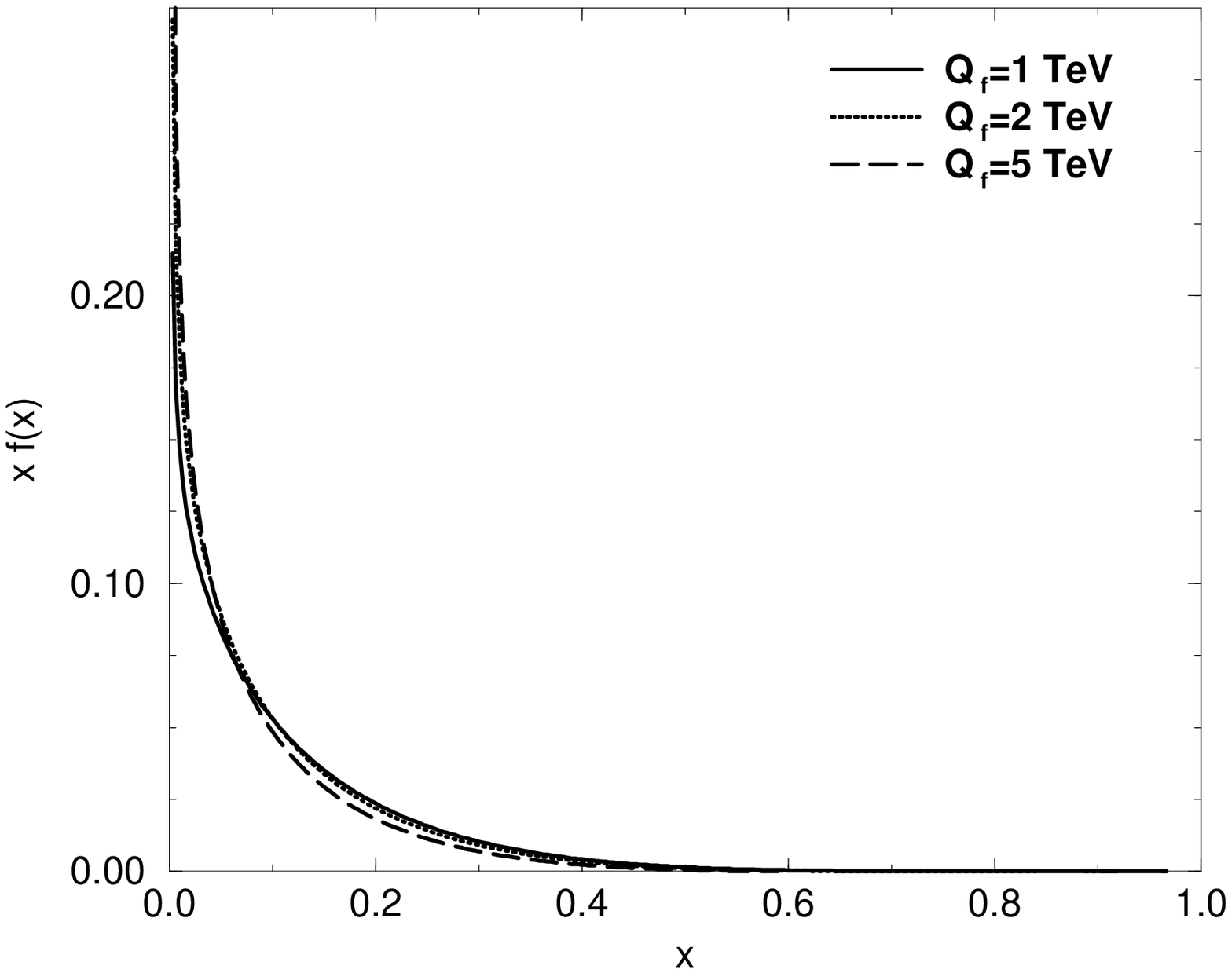}}
\caption{Gluino distribution for $Q_f=1,2$ and 5 TeV and $m_{\lambda}=100$ GeV }
\end{figure}

\begin{figure}
\centerline{\includegraphics[angle=0,width=.8\textwidth]{fig9.eps}}
\caption{Gluino and gluon distributions for very large final evolution scales 
$Q_f= 5$ and 10 TeV. The gluino mass is 250 GeV }
\end{figure}

\begin{figure}
\centerline{\includegraphics[angle=0,width=.8\textwidth]{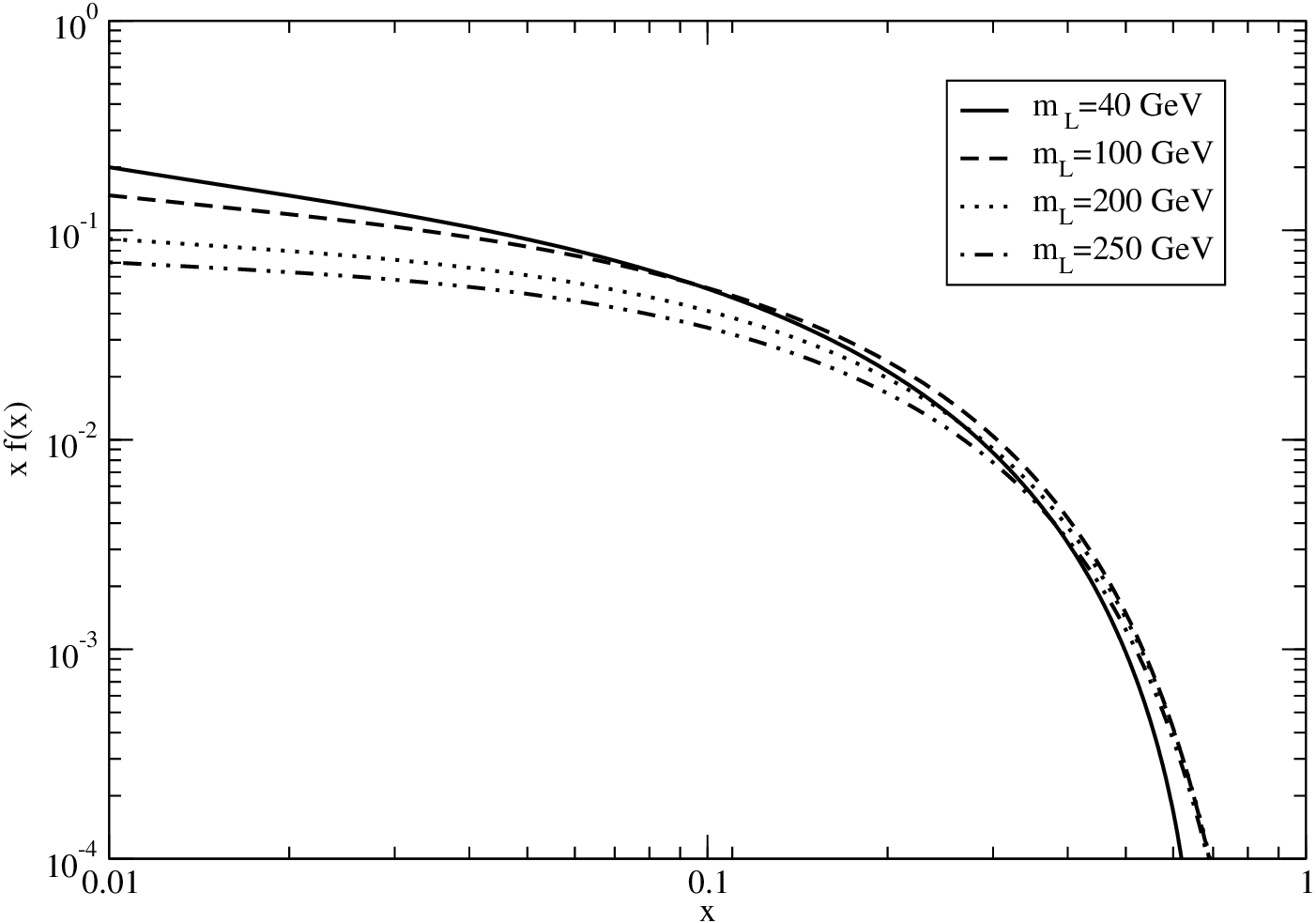}}
\caption{ Gluino distributions for a varying $m_{\lambda}$ 
(40, 100, 200 and 250 GeV) with a fixed final evolution scale $Q_f=1$ TeV.}
\end{figure}

\begin{figure}
\centerline{\includegraphics[angle=0,width=.8\textwidth]{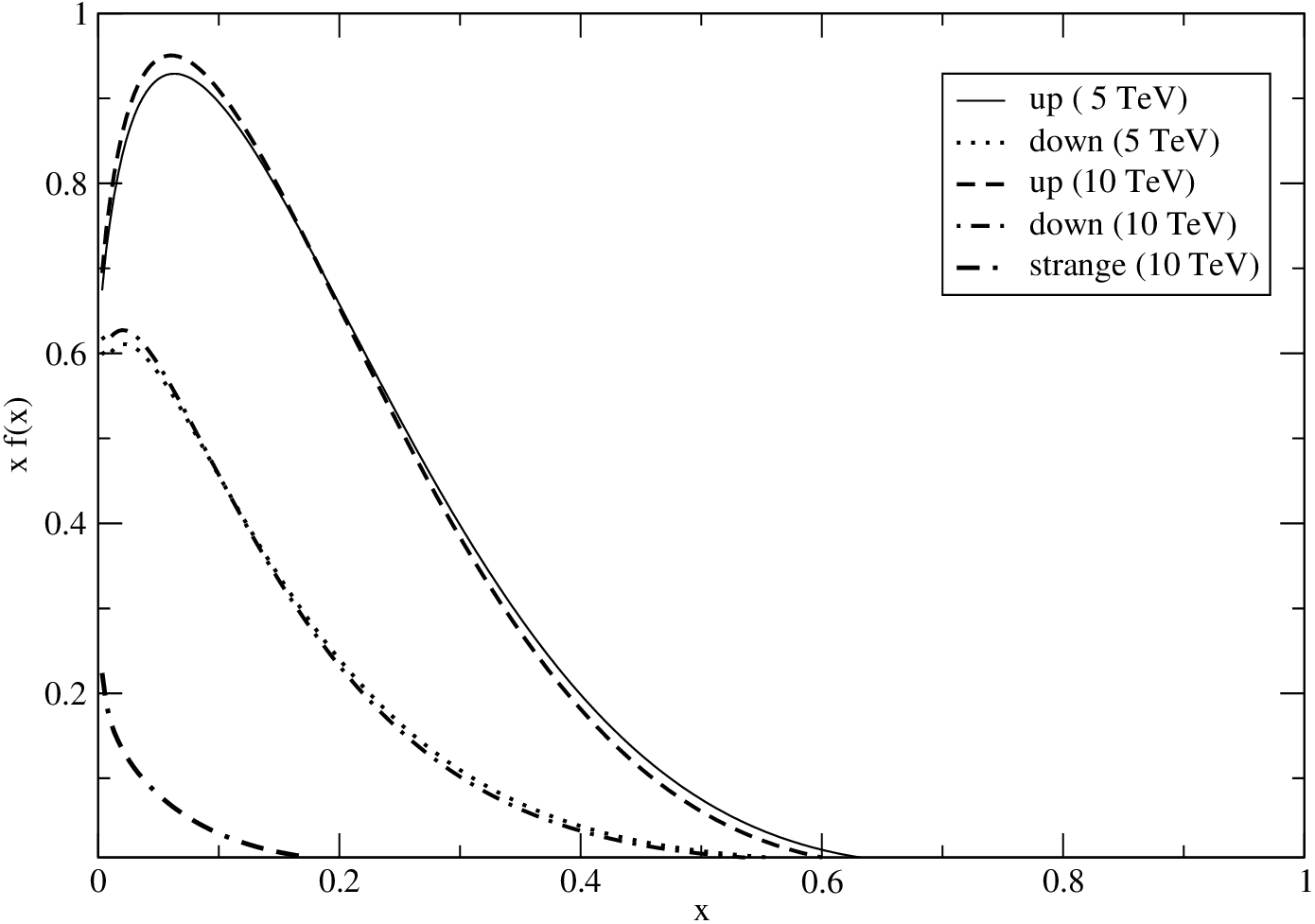}}
\caption{ u, d quark distributions $m_{\lambda}=250$ GeV 
with a varying final evolution scale $Q_f=5$ and 10 TeV. Shown is also the strange quark distribution 
with $Q_f=5$ TeV.}
\end{figure}

\begin{figure}
\centerline{\includegraphics[angle=0,width=.9\textwidth]{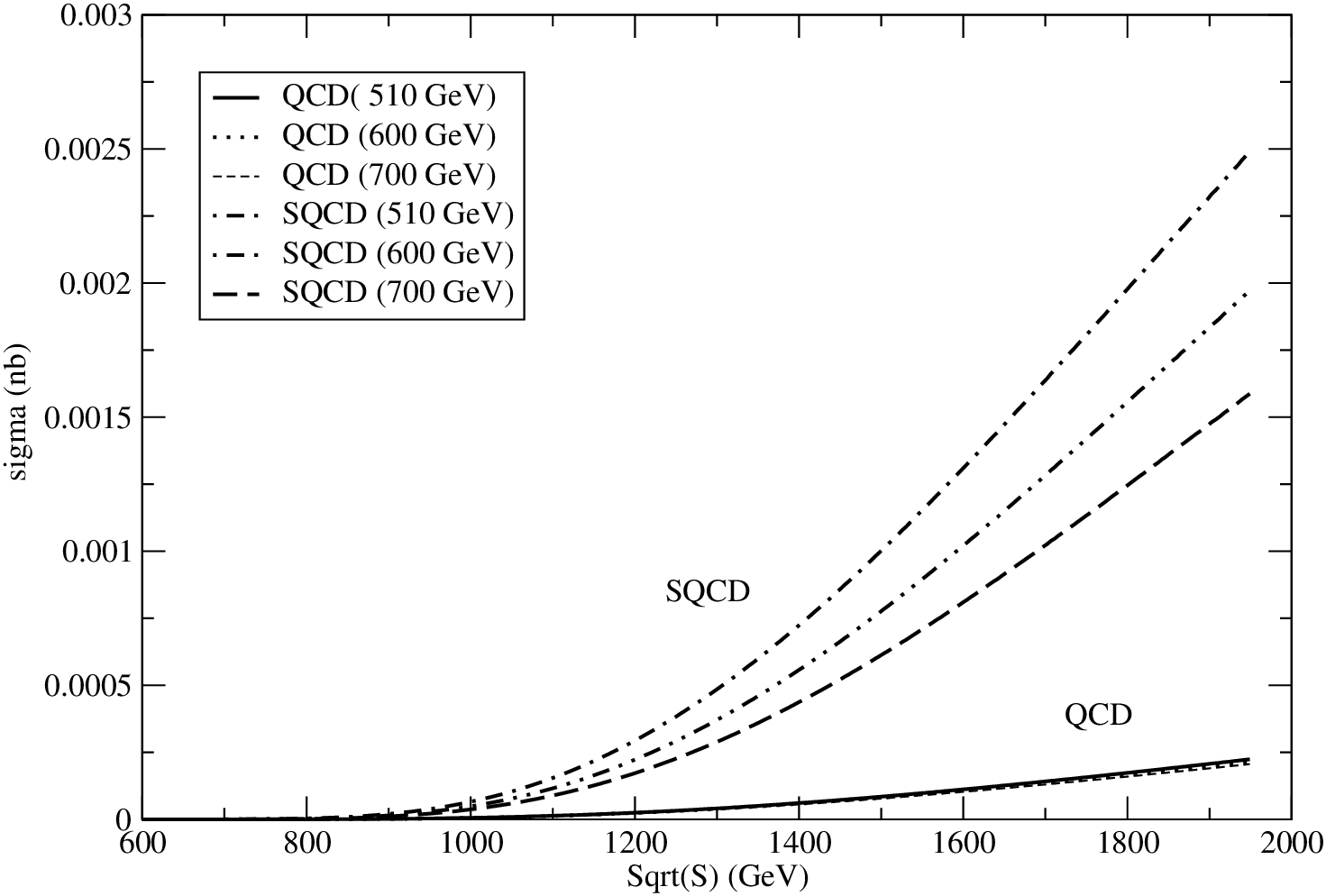}}
\caption{Dependence of the total 2-gluino cross section $\sigma_{p p \to \lambda \lambda}$
on the factorization scale in the QCD and SQCD 
cases ($m_{\lambda}=250$ GeV). Shown are the factorization scales $Q_{fact}$= 510, 600 and 700 GeV. We show both the QCD and the SQCD results }
\end{figure}

\begin{figure}
\centerline{\includegraphics[angle=0,width=.9\textwidth]{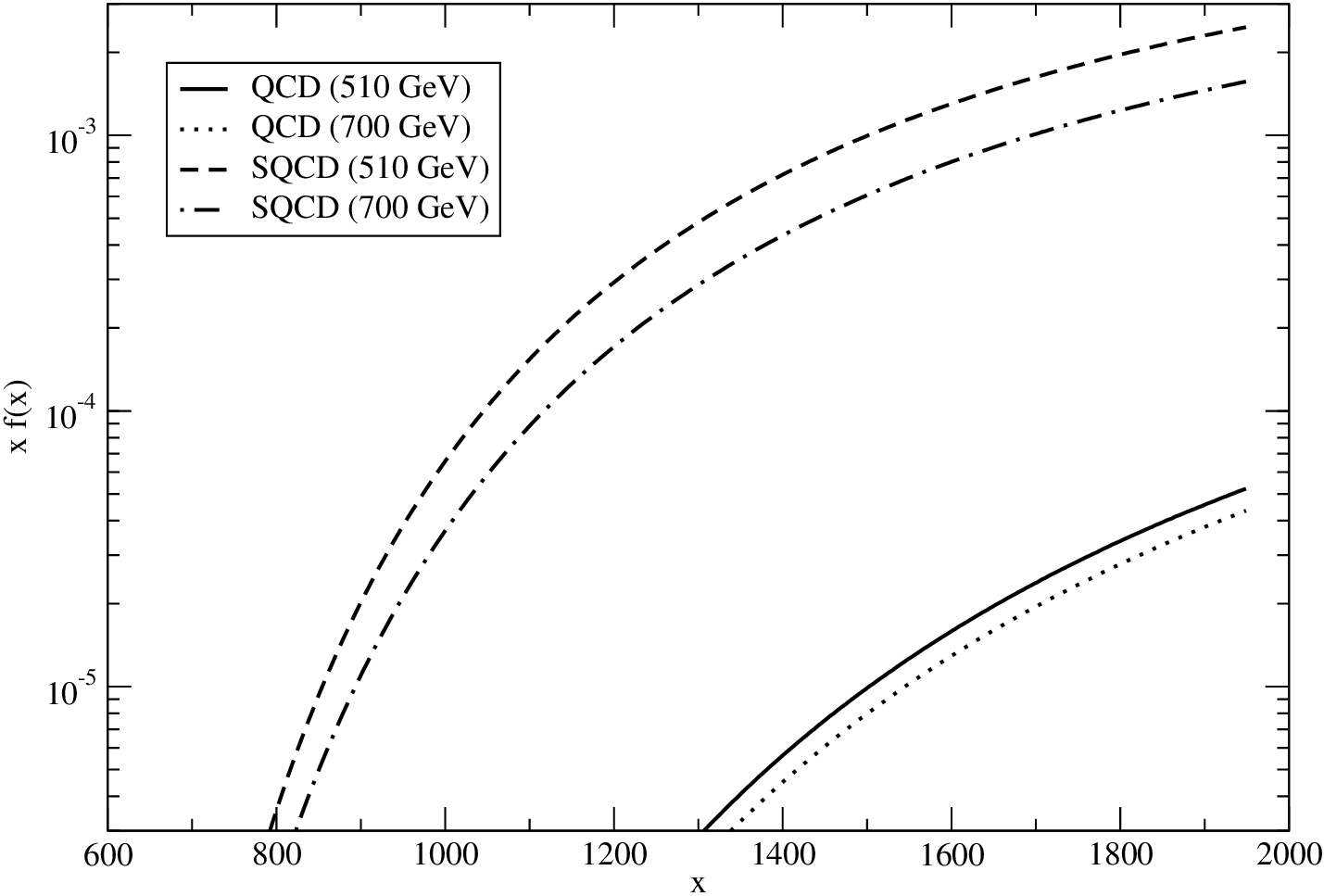}}
\caption{Dependence of the total 2-gluino cross section $\sigma_{g g \to \lambda \lambda}$
on the factorization scale in the QCD and SQCD 
cases ($m_{\lambda}=250$ GeV). Shown are the factorization scales $Q_{fact}$= 510 and 700 GeV.We show both the QCD and the SQCD results.  }
\end{figure}

\begin{figure}
\centerline{\includegraphics[angle=0,width=.9\textwidth]{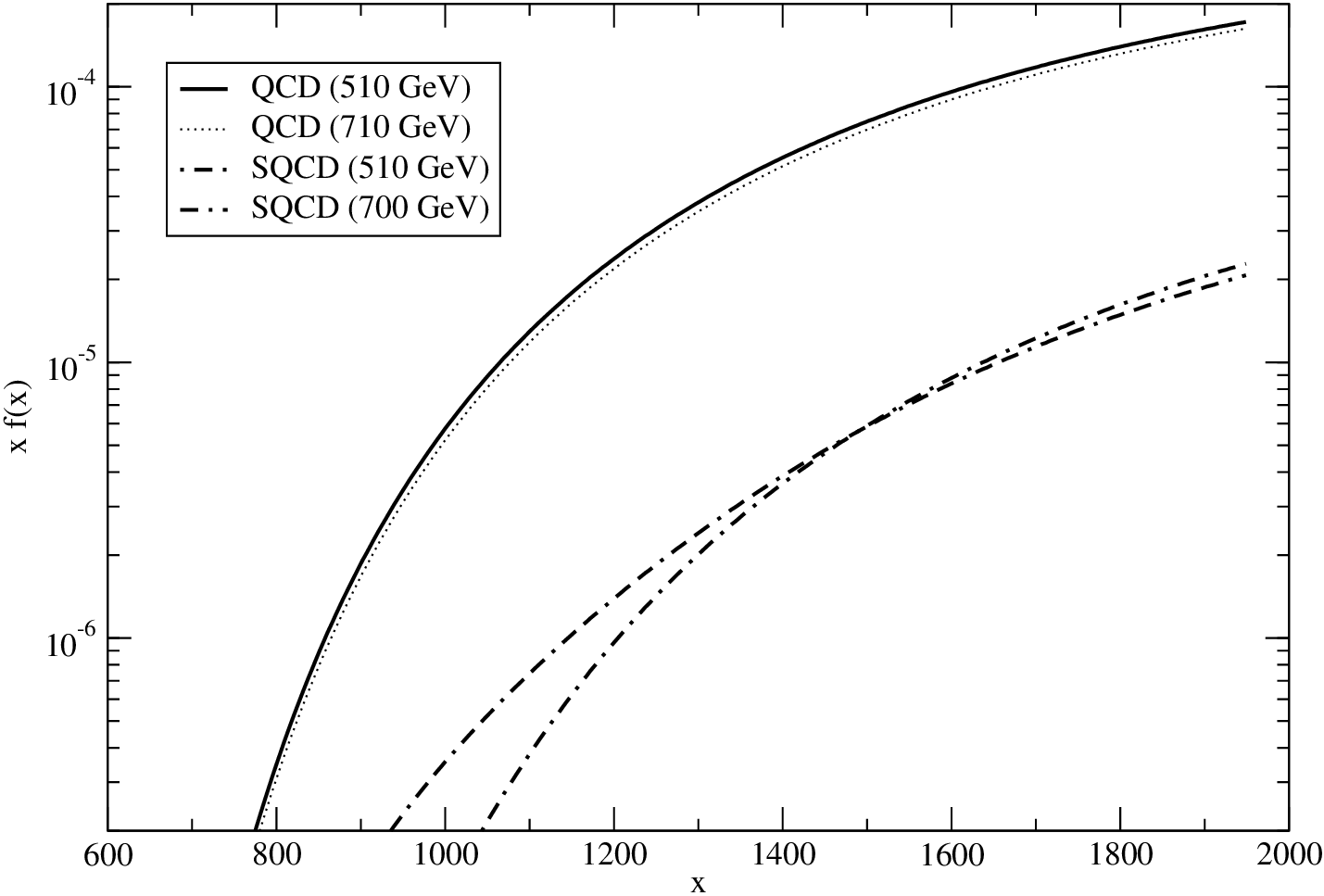}}
\caption{Dependence of the total 2-gluino cross section $\sigma_{q \bar{q} \to \lambda \lambda}$
on the factorization scale in the QCD and SQCD 
cases ($m_{\lambda}=250$ GeV). Shown are the factorization scales $Q_{fact}$= 510 and 700 GeV.We show both the QCD and the SQCD results.  }
\end{figure}

\end{document}